\documentclass[
  nofootinbib,
  11pt,
  eqsecnum,
  showkeys
]{revtex4-2}

\usepackage{geometry}

\usepackage[utf8]{inputenc}
\usepackage[T1]{fontenc}

\usepackage[english]{babel}
\usepackage{csquotes}

\usepackage{microtype}

\usepackage{mathtools}
\usepackage{amsfonts}
\usepackage{bm}
\usepackage{tensor}

\newcommand\odv[2]{\frac{\dd #1}{\dd #2}}
\newcommand\pdv[2]{\frac{\partial #1}{\partial #2}}
\newcommand\fdv[2]{\frac{\delta #1}{\delta #2}}

\usepackage{hyperref}
\hypersetup{hidelinks}
\usepackage[capitalise]{cleveref}
\newcommand\Eqref[1]{Equation \eqref{#1}}

\usepackage{graphicx}

\usepackage{suffix}
\usepackage{xifthen}

\usepackage{tikz}
\usetikzlibrary{cd}

\newcommand\blank{\cdot}

\DeclareMathAlphabet{\mathup}{OT1}{\familydefault}{m}{n}

\newcommand\mathsubscript[1]{\mathup{#1}}

\def\imagi{\mathup{i}}
\def\expe{\mathup{e}}

\renewcommand\Re{\operatorname{Re}}
\renewcommand\Im{\operatorname{Im}}
\newcommand\sign{\operatorname{sgn}}

\newcommand\mathcomma{\,,}
\newcommand\mathperiod{\,.}

\newcommand\multindex[1]{\underline{#1}}

\newcommand\evalat[1]{#1 \rvert}
\WithSuffix\newcommand\evalat*[1]{\left. #1 \right\rvert}

\newcommand\field[1]{\mathbb{#1}}

\renewcommand\vec[1]{\bm{#1}}

\def\dd{\mathrm{d}}
\newcommand\intmeasure[2][]{\dd^{#1}#2\,}
\newcommand\fourierintmeasure[2][]{\frac{\dd^{#1}#2}{(2\pi)^{#1}}\,}

\newcommand\laplace[1][]{\triangle_{#1}}

\newcommand\abs[1]{\lvert #1 \rvert}
\WithSuffix\newcommand\abs*[1]{\left\lvert #1 \right\rvert}
\newcommand\norm[1]{\lVert #1 \rVert}
\WithSuffix\newcommand\norm*[1]{\left\lVert #1 \right\rVert}
\newcommand\compconj[1]{\bar{#1}}
\WithSuffix\newcommand\compconj*[1]{\overline{#1}}

\newcommand\bigO[1]{O(#1)}
\WithSuffix\newcommand\bigO*[1]{O\left(#1\right)}

\newcommand\poissonbracket[2]{\{#1, #2\}}
\WithSuffix\newcommand\poissonbracket*[2]{\left\{#1, #2\right\}}
\newcommand\commutator[2]{[#1,#2]}
\WithSuffix\newcommand\commutator*[2]{\left[#1,#2\right]}
\newcommand\op[1]{\hat{#1}}
\WithSuffix\newcommand\op*[1]{\widehat{#1}}

\newcommand\diracdelta[1]{\delta(#1)}

\newcommand\critical{\mathsubscript{c}}

\newcommand\planck{\mathsubscript{P}}

\newcommand\liebracket[2]{[#1,#2]}
\WithSuffix\newcommand\liebracket*[2]{\left[#1,#2\right]}
\newcommand\liegroup[1]{\ensuremath{#1}}

\newcommand\ket[1]{| #1 \rangle}
\WithSuffix\newcommand\ket*[1]{\left| #1 \right\rangle}
\newcommand\bra[1]{\langle #1 |}

\newcommand\expval[2][]{\langle #2 \rangle_{#1}}
\WithSuffix\newcommand\expval*[2][]{\left\langle #2 \right\rangle_{#1}}

\newcommand\hermconj[1]{#1^\dagger}

\newcommand\normalorder[1]{: #1 :}

\newcommand\vecgrad{\vec{\nabla}}

\newcommand\gftkcoeff[2][J]{\mathcal{K}_{#1}^{(#2)}}

\newcommand\singlemode{\text{sm}}
\newcommand\doublemode{\text{dm}}

\newcommand\threequantity[1]{{}^{(3)}\!#1}
\newcommand\threecurvature{\threequantity{R}}
 \begin{document}
  \title{Hamiltonian group field theory with multiple scalar matter fields}

\author{Steffen Gielen}
\email{s.c.gielen@sheffield.ac.uk}
\author{Axel Polaczek}
\email{apolaczek1@sheffield.ac.uk}
\affiliation{School of Mathematics and Statistics, University of Sheffield, Hicks Building, Hounsfield Road, Sheffield S3 7RH, United Kingdom}

\date{\today}

\begin{abstract}
One approach to defining dynamics for quantum gravity in a naturally timeless setting is to select a suitable matter degree of freedom as a `clock' before quantisation.
This idea of deparametrisation was recently introduced in group field theory leading to a Hamiltonian formulation in which states or operators evolve with respect to the clock given by a free massless scalar field, similar to what happens in deparametrised models in quantum cosmology.
Here we extend the construction of Hamiltonian group field theory to models with multiple scalar matter fields, encountering new features and technical subtleties compared with the previously studied case.
We show that the effective cosmological dynamics for these more general models reduce to the Friedmann dynamics of general relativity with multiple scalar fields in the limit of large volume, if suitable (non-generic) initial conditions are chosen.
At high energy, we find corrections to the classical Friedmann equations whose form is similar to what is found in loop quantum cosmology.
These corrections lead to generic singularity resolution by a bounce.
For generic initial conditions, the effective cosmological dynamics treat the `clock' field and other matter fields differently, in disagreement with the Friedmann dynamics of general relativity.
We speculate on a possible interpretation in terms of inhomogeneities.
\end{abstract}


\maketitle
   \tableofcontents
  \section{Introduction}

Group field theory (GFT) is a formalism in which the problem of quantum gravity can be studied from different angles and with a variety of methods \cite{Fre05,Ori06,Kra11,Ori12}.
The GFT formalism generalises the basic ideas of matrix and tensor models for quantum gravity \cite{Gin92,Gur11,Car15} by incorporating additional structure from loop quantum gravity (LQG) and spin foam models \cite{Ash04,Per12}.
A GFT is defined by an action for a `group field' living on an abstract group manifold (as opposed to on a spacetime manifold): the perturbative expansion of the GFT partition function generates a sum over Feynman amplitudes, each of which can be interpreted as a discrete (simplicial) quantum gravity amplitude.
The theory is then to be defined in the continuum limit of such a sum.
Thus, the twist that GFT adds to usual matrix and tensor models is that the purely combinatorial graph structures appearing in these models are equipped with group-theoretical data which can be interpreted as local excitations of gravity and matter, as they are in LQG and spin foam models \cite{Ori17}.

At this level, the purpose of using the GFT formalism could be seen as providing a generating function for a sum over spin foam amplitudes.
Indeed, there is a one-to-one correspondence between a wide class of spin foam amplitudes and GFT actions \cite{PFKR00,RR01}.
This correspondence then motivates viewing GFT itself as a proposal for a theory of quantum gravity.
Its Feynman expansion provides one, but not the only way of studying this theory, since any quantum field theory is defined beyond its perturbative expansion.
In particular one can study the renormalisation of GFT models, which has been done over the last years \cite{Car13,Ben14,Car16}.
Identifying a physically interesting class of renormalisable models would be key towards understanding the continuum limit of GFT, needed to make sense of any fundamentally discrete setting for quantum gravity.

For known models of interest for quantum gravity, it is hard to gain computational control over the perturbative expansion, while truncating the expansion after a few terms would not appear to be justifiable.
New methods are then needed to understand the nonperturbative dynamics of a GFT.
One proposal, related to the search for fixed points under renormalisation flow that describe continuum geometry \cite{PT20}, is that a physically relevant continuum phase can be described by a `condensate' in which the group field acquires a nonvanishing expectation value \cite{GOS14,GS16,Ori17a}.
GFT condensates have mostly been studied in a canonical approach in which one works with a Fock space generated by creation and annihilation operators obtained from the group field.
This GFT Fock space can be interpreted in terms of LQG spin networks in that the degrees of freedom carried by Fock states are the degrees of freedom of spin networks, although there are important differences in the structure of the (kinematical) Hilbert space: the Fock space has no analogue of `cylindrical consistency' \cite{Dit12} and does not distinguish between spin networks with the same number of vertices (see \cite{Ori13} for details).
The correspondence is more direct at the level of a fixed graph, which is where many studies in LQG and spin foam models take place; the state space for LQG on a fixed graph can be identified with a subspace of the GFT Fock space.

One of the main goals of this paper is to extend the canonical quantisation of GFT to a new class of models, namely, models for quantum gravity coupled to an arbitrary number $D\ge 2$ of free massless scalar fields.
While most models are proposals for pure (vacuum) quantum gravity, models with a single scalar matter field have been studied in various papers previously \cite{OSW16,OSW17,Li17}.
There are various motivations for including matter into the GFT formalism.
First of all, one might argue that matter is essential for the physical interpretation of any theory of quantum gravity; whether there is a realistic coupling of matter and gravity is one of the most basic consistency questions for any approach.
The main motivation for including scalar matter fields has come from the cosmological application of GFT \cite{PS19,Ori17a,GB20}.
Just as in traditional quantum cosmology \cite{BI75} and in loop quantum cosmology (LQC) \cite{Boj01,APS06,AS11}, in GFT a free scalar field can serve as a relational `clock' which parametrises the evolution of the spacetime geometry.
In this setting one can then derive effective Friedmann equations capturing the relational evolution of quantities like the expectation value of the total (spatial) volume relative to the matter scalar field.
Remarkably, previous work has shown that the effective dynamics of GFT condensates can reproduce the main features of LQC, reduce to the dynamics of general relativity at large volume and resolve the classical Big Bang singularity by a quantum bounce at high energies \cite{OSW16,OSW17,Gie16}.
These results have been extended in various directions \cite{CPS16,PS17,AK20}.

So far, there have been two basic approaches towards defining the canonical quantisation of a GFT.
The initial approach was to postulate commutation relations between basic field operators, and hence the creation and annihilation operators derived from them, which lead to a Fock space structure similar to the spin network Hilbert space of LQG \cite{GOS14,Ori13}.
This construction is purely kinematical and does not use the GFT dynamics or action; it leads to a kinematical Hilbert space that is too large and subject to additional constraints, just as the Hilbert space of LQG.
While this construction is close to what happens in canonical LQG, one could alternatively proceed more closely to standard quantum field theory and derive the canonical Hilbert space structure from an action.
This idea was first proposed in \cite{AGW18} in a toy model for GFT: a Fock space based on creation and annihilation operators was derived from a postulated action, with dynamics given by a `squeezing' Hamiltonian.
Time evolution was defined to be evolution with respect to the matter scalar field variable, in what is usually called a \emph{deparametrised} approach.
Rather remarkably, the main features of GFT cosmology could be reproduced in this very simple model.
The connection of this toy model to full GFT was then clarified in \cite{WEw19} where a deparametrised approach was adopted for GFT by choosing the scalar field argument as a time variable from the beginning, and the Legendre transform was computed leading to a Hamiltonian for GFT.
In the truncation to quadratic terms the resulting dynamics were very similar to that of \cite{AGW18}, again leading to LQC-like cosmological dynamics (see also \cite{GP19} for more details of the effective cosmology of this model).
The method of deparametrisation using scalar fields had previously appeared in canonical LQG \cite{GT07,DGKL10,HP11}, where it also leads to a different Hilbert space which is not subject to (Hamiltonian or diffeomorphism) constraints.
An extended discussion of these two avenues to defining a canonical quantisation and relational time evolution in GFT was recently given in \cite{MO20}.

In this paper we extend the deparametrised canonical formalism for GFT to multiple scalar matter fields.
There are multiple motivations for doing this.
By introducing a model with multiple scalar field `coordinates' the Hamiltonian methods for GFT developed in \cite{WEw19} can be extended from what is effectively quantum mechanics (a quantum system evolving in time) to quantum field theory (a field configuration on a multi-dimensional manifold), thus strengthening and generalising their foundations.
Indeed we will see that features of canonical quantisation of quantum field theories such as the need for regularisation of formally divergent quantities now appear in GFT.
Perhaps more importantly, the cosmological application of models with a single scalar field provided some of the most important results obtained in the GFT formalism overall by allowing the derivation of effective Friedmann equations with the correct physical properties.
Extending this formalism to multiple matter fields provides an important further test of the physical validity of GFT while also bringing the resulting cosmological models closer to being more realistic.

A further motivation comes from the problem of time in quantum gravity \cite{Ish92,Kuc91,And12}: deparametrised models are in danger of violating the general covariance  of general relativity since a preferred time coordinate has been chosen before quantisation.
Usually the problem of time is formulated for theories defined on a continuum spacetime, where diffeomorphisms act as gauge transformations on all dynamical fields, so that no particular time coordinate should be singled out as characterising dynamics and evolution.
The very large symmetry group of diffeomorphisms must then be implemented properly in the quantum theory.
In a fundamentally discrete setting such as GFT, the notion of diffeomorphism symmetry is less obvious: without a continuum spacetime there seems to be no room for defining coordinate transformations.
However, there are situations in which a discrete version of diffeomorphisms can be understood.
Typical examples are topological field theories where all physical solutions are required to be flat or perhaps of constant curvature, as is the case for general relativity with or without cosmological constant in three dimensions.
In such a setting, there exist discrete symmetries relating discretisations describing the same configuration, which can be interpreted as a discrete analogue of diffeomorphisms.
These can be defined, e.g., in GFT in three dimensions \cite{BGO11} and for topological spin foam models \cite{BD13,ADGRT19}.
Much less is known about discrete diffeomorphisms in models for quantum gravity with local degrees of freedom but the usual notion of diffeomorphism symmetry should emerge in a continuum limit of such theories \cite{Ditt12}, including GFT; it is hence an important issue to investigate.

Studying a GFT with multiple matter fields allows us to study, for the first time, the covariance of Hamiltonian GFT with respect to certain global coordinate changes:
given that we use a matter degree of freedom as a relational clock, such a model contains multiple candidates for matter clocks, and we can study the behaviour of the theory under passing from one clock to another.
As we will see, one of the more striking features of the Hamiltonian GFT formalism with multiple scalar fields is that it is not covariant under the full symmetry group of the GFT action, which includes arbitrary translations or rotations among any of the matter fields (which all appear in the action on the same footing).
These symmetries are global symmetries in GFT, corresponding to linear (global) transformations between different choices of clock.
They are the analogue of the Poincar\'e transformations of special relativity or a field theory defined on Minkowski spacetime.
We find that in Hamiltonian GFT this symmetry is broken to translations and rotations that involve only the matter fields that have not been selected as a clock, whereas the clock field is treated separately.
This symmetry breaking will appear at various points in our analysis -- it can be understood already at the classical level as arising from treating an elliptic partial differential equation as an initial value problem, which singles out a preferred direction.
This result has a direct impact on the resulting effective cosmological dynamics, which likewise treats the clock matter field differently from all other matter fields.
Given that these global transformations should correspond to particular types of coordinate transformations in a continuum limit (similar to how global Lorentz transformations are particular diffeomorphisms on Minkowski spacetime), we expect these findings to be relevant for the more general question of whether a Hamiltonian formalism for GFT can preserve general covariance (i.e., covariance with respect to arbitrary coordinate changes).

The structure of the paper is as follows.
\Cref{sec:multi_scalars} introduces the general formalism for GFT coupled to $D\ge 2$ free massless scalar fields, and uses symmetry arguments and a derivative expansion to constrain the possible form of the (quadratic) GFT actions.
\cref{sec:canonical_quant} details the usual steps of canonical quantisation, defining a canonical momentum to the group field and introducing creation and annihilation operators.
We identify conserved quantities from symmetries of the action, mention observables of interest, and define Fock coherent states.
In \cref{sec:eff_cosmology} we derive the effective cosmological dynamics given in terms of effective Friedmann equations for the total volume.
The general case depends on infinitely many initial conditions, and we then specify to cases in which only one or two field modes are relevant.
These dynamics can match those of general relativity at late times, but this requires relations between initial conditions even for simple coherent states.
We do find generic resolution of the big bang singularity by a bounce, with high-energy corrections very similar to those of LQC.
We finish with a discussion of our results in \cref{sec:discussion}.

   \section{GFT for quantum gravity with multiple scalar fields}
\label{sec:multi_scalars}

GFT models that are proposals for a theory of quantum gravity are usually defined in such a way that the perturbative expansion of their partition function can be interpreted as a sum over discrete quantum gravity amplitudes.
This requirement informs the choice of fields appearing in the theory as well as the choice of dynamics or action for these fields.

We focus on models for which the discrete structures appearing in the perturbative expansion can be interpreted as simplicial geometries, that is, extended structures constructed by gluing a number of simplicial `building blocks' by specified rules.
These  structures carry the degrees of freedom of discrete geometries appearing in loop quantum gravity and spin foams: group elements representing discrete parallel transports of a gravitational connection from one building block to the next and, in the case of interest here, scalar matter degrees of freedom associated to the building blocks themselves.

The Feynman graphs in the GFT expansion represent \emph{spin foams},  2-complexes in which each face is labelled by an irreducible representation of a Lie group $G$ and each edge is labelled by an intertwiner between representations.
In the case where scalar matter is present, edges also carry labels corresponding to the values of these matter fields.
GFT boundary data live on \emph{spin networks} containing the edges and vertices on the boundary of a spin foam.
Since edges bound faces and vertices bound edges, a spin network carries representation labels on its edges and intertwiner/matter labels on its vertices.
An excellent introduction to spin networks and spin foams for quantum gravity is given in \cite{Bae99}.

One way of defining the group field is to consider the kind of boundary states, or spin networks, one wants to generate.
First of all, this involves a choice of combinatorics: for models for four-dimensional quantum gravity, the most common choice is to restrict these spin networks to be four-valent so that, in the dual interpretation, boundary states represent simplicial geometries built out of tetrahedra.
This restriction to simplicial geometries is also often made in spin foam models.
Spin foam models have been generalised to arbitrary valency \cite{KKL09} and this can also be done for GFT \cite{ORT14}, but we will not do so here.

Adopting this common choice of modelling four-dimensional quantum gravity using four-valent spin networks as the boundary states, we consider a real field
\begin{equation}
  \varphi:
  G^4 \times \field{R}^D
  \rightarrow
  \field{R}
  \mathcomma
  \quad
  (g_a, \chi^\alpha)
  \mapsto
  \varphi(g_a, \chi^\alpha)
\end{equation}
and write the most general GFT action for such a field as
\begin{equation}
  \label{eq:action_general}
  S[\varphi]
  =
  \mathcal{K}[\varphi]
  +
  \mathcal{V}[\varphi]
  \mathcomma
\end{equation}
where $\mathcal{K}[\varphi]$ contains terms bilinear in $\varphi$ and $\mathcal{V}[\varphi]$ contains terms of higher order, i.e., trilinear and higher.
Notice that, since we restricted the spin networks representing GFT states to valency four and each spin network edge is associated with a $G$ representation label, the field $\varphi$ has four $G$-valued arguments $g_a$.
The $\field{R}$-valued arguments $\chi^\alpha$ represent matter scalar fields; we assume there are $D$ of these, where $D\ge 2$.
The models we consider then extend GFT models with a single matter field, such as the ones in \cite{WEw19}.
In the following we will also restrict ourselves to $G=\liegroup{SU}(2)$, although other choices would be possible.

In the perturbative expansion of such a theory, the `kinetic' part $\mathcal{K}[\varphi]$ encodes the propagator and the higher-order part $\mathcal{V}[\varphi]$ generates interaction vertices which are responsible for the gluing of elementary building blocks to form higher-dimensional structures.
The type of discrete geometries appearing in the perturbative expansion depends sensitively on $\mathcal{V}[\varphi]$.
For example, to generate four-dimensional simplicial geometries the interaction terms should involve fifth powers of $\varphi$ and include the combinatorial pattern corresponding to the gluing of five tetrahedra to a 4-simplex, as in the topological Ooguri model \cite{Oog92}.

In this paper we will restrict our attention to the free theory which only includes $\mathcal{K}[\varphi]$ in order to construct a Hamiltonian formalism.
This is as in usual quantum field theory where canonical quantisation is based on a quadratic action and interactions are then added perturbatively.
We expand the group field in modes associated to $\liegroup{SU}(2)$ representations,
\begin{equation}
  \label{eq:pw_decomp}
  \varphi  (g_a, \chi^\alpha)
  =
  \sum_J
  \varphi_J (\chi^\alpha)
  D_J(g_a)
\end{equation}
following the economical notation of \cite{AK20}: assuming the usual `gauge invariance' condition
\begin{equation}
  \label{eq:gauge_inv}
  \varphi(g_a,\chi^\alpha)
  =
  \varphi(g_ah,\chi^\alpha)
  \quad
  \forall\;
  h\in \liegroup{SU}(2)
  \mathcomma
\end{equation}
a basis of square-integrable functions on $\liegroup{SU}(2)^4$ satisfying \eqref{eq:gauge_inv} is given by
\begin{equation}
  \label{eq:pw_basis}
  D_J(g_a)
  =
  \sum_{\multindex{n}}\mathcal{I}^{\multindex{j}, \iota}_{\multindex{n}}
  \prod_{a=1}^4
  \sqrt{2j_a+1}\,
  D^{j_a}_{m_a n_a}(g_a)
\end{equation}
where $J = (\multindex{j},\multindex{m},\iota)$ is a multi-index including half-integer representation labels $\multindex{j}$ and magnetic indices $\multindex{m}$ and intertwiner labels $\iota$ whose possible values depend on $\multindex{j}$.
The sum is over magnetic indices $\multindex{n}$; $\mathcal{I}^{\multindex{j}, \iota}_{\multindex{n}}$ denotes the intertwiner labelled by $\iota$
\footnote{Depending on $\multindex{j}$ there may be no, one, or multiple linearly independent intertwiners, i.e., equivariant linear maps from the tensor product of representations $\multindex{j}$ to the trivial representation.
  $\iota$ runs over a basis.
}
and $D^j_{mn}(g)$ are Wigner $D$-matrices.
Note that for a real group field the modes $\varphi_J$ in \eqref{eq:pw_decomp} are not independent.
Indeed, they must satisfy the reality condition \cite{Oog92}
\begin{equation}
  \label{eq:group_field_reality_condition}
  \varphi_{-J}(\chi^\alpha)
  =
  \epsilon_J
  \compconj{\varphi}_J(\chi^\alpha)
  \mathcomma
\end{equation}
where $-J = (\multindex{j}, - \multindex{m}, \iota)$ and $\epsilon_J = (-1)^{\sum_i (j_i - m_i)}$.

We then assume that after suitable redefinition of the $J$ basis, the kinetic term is diagonal in this basis.\footnote{The `standard' kinetic term is not diagonal in the basis \eqref{eq:pw_basis}.
  Diagonalisation can then be achieved by splitting the modes $\varphi_J$ into their real and imaginary parts; see \cref{app:diagonal_kinetic} for details.
}
Explicitly, the kinetic term then takes the form
\begin{equation}
  \label{eq:kinetic_term}
  S[\varphi]
  =
  \frac{1}{2}
  \sum_J
  \int
  \intmeasure[D]{\chi}
  \intmeasure[D]{\chi'}
  \varphi_J(\chi^\alpha)
  \mathcal{K}_J(\chi^\alpha, \chi'{}^\alpha)
  \varphi_J(\chi'{}^\alpha)
  \mathperiod
\end{equation}
The form of $\mathcal{K}_J(\chi^\alpha, \chi'{}^\alpha)$, which is general at this point, can be further constrained by symmetry arguments.

\subsection{Specifying the kinetic term}
\label{sec:kinetic_term}
The construction of GFT models for quantum gravity coupled to matter is, in the general case, a difficult problem.
There is no proposal, for instance, for a GFT that would include the entire standard model of particle physics.
However, in the special case where matter is given by free massless scalar fields, the large amount of symmetry in the matter sector constrains the possible GFT actions quite significantly.
This idea was exploited in the construction of GFT models for quantum gravity with a single massless scalar field in \cite{OSW16,OSW17}; the general form of the action allowed by symmetry could then be further specified by constructing an explicit model \cite{Li17}.
The symmetry argument was extended to multiple scalar matter fields in \cite{GO18}.
Here we will be slightly more specific than \cite{GO18} in that we will assume that all $D=d+1$ matter scalar fields are (minimally) coupled to gravity in exactly the same way (rather than allowing for the case that one `clock' matter field is coupled differently from the other `rod' matter fields).
Recall that we are assuming $D\ge 2$ so that $d\ge 1$.
We will label the matter fields by indices running from $0$ to $d$.

In conventional quantum field theory, the Lagrangian density of $D=d+1$ massless scalar fields minimally coupled to an arbitrary spacetime metric $g_{\mu\nu}$ is given by
\begin{equation}
  \mathcal{L}
  =
  - \frac{1}{2}
  \sum_{\alpha = 0}^{d}
  \sqrt{-g}\,
  g^{\mu\nu}\,
  \partial_\mu
  \chi^\alpha
  \partial_\nu
  \chi^\alpha
  \mathperiod
\end{equation}
This Lagrangian is invariant under the action of the Euclidean group $\liegroup{E}(D)$ acting on the massless scalar fields via
\begin{itemize}
\item
  Translations: $\chi^\alpha \mapsto \chi^\alpha + a^\alpha$,
\item
  Rotations: $\chi^\alpha \mapsto \sum_\beta R\indices{^\alpha_\beta} \chi^\beta$,
\item
  Reflections: $\chi^\alpha \mapsto - \chi^\alpha$ (for each $\chi^\alpha$ separately),
\end{itemize}
where $a^\alpha\in \field{R}^D$ and $R\indices{^\alpha_\beta}\in\liegroup{O}(D)$  are spacetime constants.

It is natural to assume that any proposal for a theory of quantum gravity coupled to this type of matter should exhibit these symmetries as well.
If we therefore demand that \eqref{eq:kinetic_term} also has these symmetries, the kinetic kernel must take the form
\begin{equation}
  \mathcal{K}_J(\chi^\alpha, \chi'^\alpha)
  =
  \mathcal{K}_J((\chi - \chi')^2)
  \mathcomma
\end{equation}
where $(\chi - \chi')^2 = \sum_\alpha [(\chi - \chi')^\alpha]^2$.
Performing an expansion in terms of derivatives in \eqref{eq:kinetic_term} then gives the action
\begin{equation}
  \label{eq:action_derivative}
  S[\varphi]
  =
  \frac{1}{2}
  \sum_J
  \int
  \intmeasure[D]{\chi}
  \sum_{n=0}^{\infty}
  \varphi_J(\chi^\alpha)
  \mathcal{K}_J^{(2n)}
  \laplace[\chi]^{n}
  \varphi_J(\chi^\alpha)
  \mathcomma
\end{equation}
where we defined
\begin{align}
  &
  \mathcal{K}^{(2n)}_J
  =
  \int
  \intmeasure[D]{\delta\chi}
  \frac{1}{(2n)!}
  \mathcal{K}_J((\delta\chi)^2)
  ((\delta{\chi})^2)^{n}
  \mathcomma
  \\
  &
  \laplace[\chi]
  =
  \sum_{\alpha=0}^d
  \left(
    \pdv{}{\chi^\alpha}
  \right)^2
  \mathperiod
\end{align}
Because of the $\liegroup{E}(D)$ symmetry of the theory the derivative expansion only contains powers of the Laplacian $\laplace[\chi]$ on $\field{R}^D$.
\Eqref{eq:action_derivative} is then the generalisation of the expansion for GFT models with a single matter field as proposed in \cite{OSW16,OSW17}.

In general \eqref{eq:action_derivative} contains an infinite number of terms with an increasing number of derivatives.
In the spirit of effective field theory one can restrict oneself to the case in which higher than second-order derivatives are negligible and assume that the kinetic term is given by
\begin{equation}
  \label{eq:kinetic_term_truncated}
  \mathcal{K}[\varphi]
  =
  \frac{1}{2}
  \int
  \intmeasure[D]{\chi}
  \sum_J
  \varphi_J(\chi^\alpha)
  \left[
    \mathcal{K}^{(0)}_J
    +
    \mathcal{K}^{(2)}_J
    \laplace[\chi]
  \right]
  \varphi_J(\chi^\alpha)
  \mathcomma
\end{equation}
where $\gftkcoeff{0}$ and $\gftkcoeff{2}$ are constants that can be derived from the initial GFT action.
Again, this is in analogy with previous work \cite{OSW16,OSW17,WEw19}.

Note that a scalar field in four-dimensional spacetime has mass dimension $1$.
Therefore the derivative expansion in \eqref{eq:action_derivative} can be associated with an expansion in powers of the Planck mass squared $m_\planck^2$.
In the rest of the paper we will work with GFT actions of the form \eqref{eq:kinetic_term_truncated} for which we define a Hamiltonian formalism.

\subsection{Equations of motion}
\label{sec:lagrangian_eom}

The equations of motion of a theory defined by the action \eqref{eq:kinetic_term_truncated} are given by
\begin{equation}
  \label{eq:eom}
  (
    \laplace[\chi]
    -
    m_J^2
  )
  \varphi_J(\chi^\alpha)
  =
  0
  \mathcomma
\end{equation}
where we have introduced the shorthand $m_J^2 = - \gftkcoeff{0} / \gftkcoeff{2}$ as in previous work on GFT cosmology \cite{OSW16}.
In spite of the suggestive notation, $m_J^2$ is not positive definite: its sign depends on the signs of the couplings $\gftkcoeff{0}$ and $\gftkcoeff{2}$ in the GFT action under consideration.
Notice also that in natural units ($\hbar=c=1$) $m_J$ has units of length rather than mass: in the effective cosmological dynamics of GFT $m_J$ defines the effective Planck length.
On the other hand, $m_J^2$ appears in \eqref{eq:eom} on the same footing as an $m^2$ term in standard (Euclidean) scalar field theory.
It is mainly for this reason that we adopt the notation from \cite{OSW16} here.

The case $m_J^2>0$ is most relevant for cosmological applications since it allows for exponentially growing (and decaying) solutions which can be interpreted as resulting in an expanding (or collapsing) geometry rather than a `static' one formed by oscillatory modes.
Let us stress that in general each $J$ mode can have a different value of $m_J^2$, and that in general some modes will have $m_J^2>0$ and others will have $m_J^2<0$.
A concrete and common example would be a kinetic term involving Laplace--Beltrami operators on $\liegroup{SU}(2)$, for which
\begin{equation}
  \gftkcoeff{0}
  =
  \mu
  -
  \sum_a
  j_a(j_a+1)
  \mathcomma
  \quad
  \gftkcoeff{2}
  =
  \tau
  \quad
  \Rightarrow
  \quad
  m_J^2
  =
  \frac{\sum_a j_a(j_a+1)-\mu}{\tau}
  \mathperiod
\end{equation}
As pointed out in \cite{Gie16}, for this form of kinetic term and assuming $\tau<0$ and $\mu>0$, only a finite number of `low' spins $j_a$ have $m_J^2>0$ and lead to exponentially growing solutions.
We will keep the form of $m_J^2$ general in what follows.

The equation of motion \eqref{eq:eom}, for each mode $J$, is an eigenvalue equation for the Laplace operator on $\field{R}^D$.
This is an elliptic partial differential equation.
Such an equation has a well-posed boundary value problem, where the values of $\varphi_J$ would be prescribed on the boundary of a suitably chosen domain in $\field{R}^D$.
To define a Hamiltonian formalism for this GFT, we will however study an initial value problem where a particular direction in $\field{R}^D$ plays the role of `time' and initial data are specified on a hypersurface orthogonal to this direction.
Such an initial value problem is not well-posed; it is unstable in the sense that small perturbations of the initial data grow exponentially.\footnote{The Cauchy problem for a Laplace equation is probably the most famous example of an ill-posed problem in analysis \cite{TA77,Pa75}.
}
As we already noted, this instability is in fact necessary to obtain a physically satisfactory cosmology.
Nevertheless, the attempt to define an initial value problem for a differential equation of elliptic type will lead to some peculiarities and pathologies in the resulting Hamiltonian theory.
In particular, the introduction of a preferred `time' direction will force us to break the $\liegroup{E}(D)$ symmetry of the action \eqref{eq:kinetic_term_truncated} and field equation \eqref{eq:eom} to a smaller group $\liegroup{E}(d)\times\liegroup{E}(1)$, where transformations mixing the `time' and the remaining directions are no longer symmetries of the theory.

Our initial assumption that we can restrict ourselves to a free action quadratic in the group fields therefore has severe limitations.
While it may be possible to neglect interaction terms for suitable initial conditions, the exponential instability in the free theory will mean that interactions dominate after a finite time.
We therefore expect the physical picture provided by the Hamiltonian formalism to be at best a good approximation for certain initial states and finite times.
Asymptotic statements where the Hamiltonian evolution is extrapolated to infinity are not strictly valid.
They can, however, provide reliable approximations for a period of time which may be sufficiently long for practical purposes.
These are well-known issues in the study of cosmological dynamics from GFT \cite{OSW16,OSW17,GP19}.
All proposals for GFT models for quantum gravity are interacting field theories, presumably only well-defined in the path integral setting.

   \section{Canonical quantisation}
\label{sec:canonical_quant}

In this section we generalise the canonical formalism of GFT introduced in \cite{WEw19} to the case where the group field takes multiple massless scalar fields as arguments.
In the previously studied case of a single massless scalar field, each group field mode (corresponding to a specific set of Peter--Weyl representation labels) evolves only in scalar field `time', and the resulting formalism is that of standard quantum mechanics.
Inclusion of one or more additional massless scalar fields then results in a formalism analogous to quantum field theory: one scalar field plays the role of time and the other fields are analogous to spatial coordinates.\footnote{
Indeed, a sufficiently large number of scalar fields can be used to construct a relational coordinate system for space and time \cite{GO18,GOW18,Gie18}.
}
We will see in detail how the resulting Hamiltonian GFT, restricted to a single set of representation labels, is analogous to a scalar field theory in standard canonical quantisation.

\subsection{Distinguished scalar field as relational clock}
\label{sec:distinguished_field}

We now split the $D$ massless scalar fields as $\chi^\alpha = (\chi^0, \vec{\chi})$, where
$\vec{\chi} = (\chi^1, \dotsc, \chi^d)$.
It should be clear that, since all $D$ matter fields are coupled in the same way, any of them can be singled out as a distinguished field (here $\chi^0$).
The label is in line with the usual coordinate labels in spacetime physics, where $0$ denotes time: $\chi^0$ will serve as a relational clock from now on.
While we refer to the fields $\vec{\chi}$ as spatial, we will not propose a particular interpretation of these fields as coordinates for space.

We now also assume that the GFT action is specified by the effective field theory kinetic term \eqref{eq:kinetic_term_truncated}, and that the interaction term $\mathcal{V}[\varphi]$ in \eqref{eq:action_general} is negligible.
The latter assumption can be viewed, alternatively, as working at zeroth order in the GFT coupling(s) in the interaction picture.

Having identified a variable as time allows to compute the conjugate momentum to $\varphi_J$,
\begin{equation}
  \label{eq:gft_field_conjugate_momentum}
  \pi_J(\chi^\alpha)
  =
  \fdv{S[\varphi]}{\left(\partial_{\chi^0} \varphi_J(\chi^\alpha)\right)}
  =
  -
  \gftkcoeff{2}
  \pdv{\varphi_J(\chi^\alpha)}{\chi^0}
  \mathperiod
\end{equation}
The kinetic term \eqref{eq:kinetic_term_truncated} can then be written as (with field arguments omitted)
\begin{equation}
  \mathcal{K}[\varphi]
  =
  \int
  \intmeasure[D]{\chi}
  \sum_J
  \left(
    \pi_J
    \pdv{\varphi_J}{\chi^0}
    -
    \frac{\gftkcoeff{2}}{2}
    \left(
      -
      \frac{1}{\abs{\gftkcoeff{2}}^2}
      \pi_J^2
      +
      \varphi_J
      \left(
        -
        \left(\pdv{}{\vec{\chi}}\right)^2
        +
        m_J^2
      \right)
      \varphi_J
    \right)
  \right)
\end{equation}
where we introduced the notation $\left(\pdv{}{\vec{\chi}}\right)^2$ for the Laplacian on $\field{R}^d$ which only acts on the `spatial' matter fields.
This expression allows one to read off the free Hamiltonian
\begin{equation}
  \label{eq:hamiltonian_sum}
  H
  =
  \int
  \intmeasure[d]{\vec{\chi}}
  \sum_J
  \frac{\gftkcoeff{2}}{2}
  \left(
    -
    \frac{1}{\abs{\gftkcoeff{2}}^2}
    \pi_J(\chi^\alpha)^2
    +
    \varphi_J(\chi^\alpha)
    \left(
      -
      \left(\pdv{}{\vec{\chi}}\right)^2
      +
      m_J^2
    \right)
    \varphi_J(\chi^\alpha)
  \right)
  \mathperiod
\end{equation}

Decomposing the fields in their Fourier modes on $\field{R}^d$
\begin{subequations}
  \begin{align}
    &
    \varphi_J(\chi^\alpha)
    =
    \int
    \fourierintmeasure[d]{\vec{k}}
    \expe^{\imagi \vec{k} \cdot \vec{\chi}}
    \varphi_J(\chi^0, \vec{k})
    \mathcomma
    \\
    &
    \pi_J(\chi^\alpha)
    =
    \int
    \fourierintmeasure[d]{\vec{k}}
    \expe^{\imagi \vec{k} \cdot \vec{\chi}}
    \pi_J(\chi^0, \vec{k})
  \end{align}
\end{subequations}
allows one to write $H = \sum_J \int \fourierintmeasure[d]{\vec{k}} H_J(\vec{k})$, with the definition of the single-mode Hamiltonian (density)
\begin{equation}
  \label{eq:hamiltonian}
  H_J(\vec{k})
  =
  \frac{\gftkcoeff{2}}{2}
  \left(
    -
    \frac{1}{\abs{\gftkcoeff{2}}^2}
    \pi_J(\chi^0, - \vec{k})
    \pi_J(\chi^0, \vec{k})
    +
    \omega_J(\vec{k})^2
    \varphi_J(\chi^0, - \vec{k})
    \varphi_J(\chi^0, \vec{k})
  \right)
  \mathcomma
\end{equation}
where we defined $\omega_J(\vec{k})^2 = \vec{k}^2 + m_J^2$.
Clearly at the level of the free theory all $J$ modes and all wavenumbers $\vec{k}$ are decoupled and can be studied separately.

\begin{figure}[htbp]
  \centering
  \includegraphics{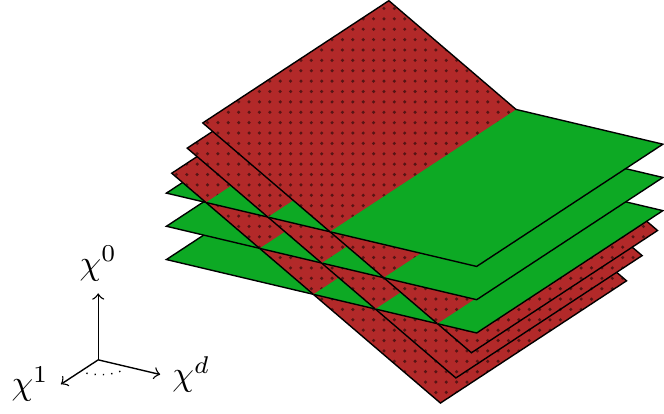}
  \caption{The formalism presented in the main text requires a particular foliation of $\field{R}^D$, namely in terms of $\chi^0 = \text{const}$ hyperplanes which are shown in green (plain).
    The group field and its momentum will be regular on each leaf only for this foliation.
    Any other choice, exemplified by the foliation shown in red (dotted), would lead to nonregular field configurations on each leaf.
  }
  \label{fig:foliation}
\end{figure}

The use of a Fourier transform to decompose the fields into wavenumbers is familiar and may appear innocent, but in a sense indicates the key feature of the Hamiltonian formalism for GFT that we will encounter in this paper.
Assuming that the fields $\varphi_J$ and $\pi_J$, seen as functions of the $\vec{\chi}$ at a given value of $\chi^0$, can be expanded into Fourier modes means that we are excluding from this point modes which grow exponentially in any of the directions parametrised by $\vec{\chi}$.
Such restrictions are needed to define an initial value problem: initial data are not arbitrary but satisfy certain regularity conditions.
Here we might require that the fields $\varphi_J$ and $\pi_J$ are square-integrable on each `constant time' hyperplane $\field{R}^d$.
This is illustrated in \cref{fig:foliation}.
Such a requirement would be particularly natural when the purpose of group field theories is seen as a second quantisation of LQG spin networks \cite{Ori13} whose wavefunctions similarly have to be square-integrable on the configuration space of a given spin network.
Nevertheless such a requirement breaks the $\liegroup{E}(D)$ covariance of the Lagrangian theory.
We have introduced a preferred slicing of $\field{R}^D$ into constant $\chi^0$ hyperplanes and demand that fields are regular on each hyperplane, whereas generic solutions grow exponentially in the $\chi^0$ direction and hence would not satisfy square integrability with respect to any other slicing of $\field{R}^D$.

\subsection{Quantisation}
In the Hamiltonian framework it is straightforward to pass to a quantum theory.
In this case we promote the group field and its conjugate momentum to operators that satisfy the equal-time canonical commutation relations
\begin{equation}
  \commutator{
    \op{\varphi}_J(\chi^0, \vec{\chi})
  }{
  \op{\pi}_{J'}(\chi^0, \vec{\chi}')
  }
  =
  \imagi
  \delta_{JJ'}
  \diracdelta{\vec{\chi} - \vec{\chi}'}
\end{equation}
which can alternatively be stated in terms of the Fourier modes
\begin{equation}
  \commutator{
    \op{\varphi}_J(\chi^0, \vec{k})
  }{
  \op{\pi}_{J'}(\chi^0, \vec{k}')
  }
  =
  \imagi
  \delta_{JJ'}
  (2\pi)^d
  \diracdelta{\vec{k} + \vec{k}'}
  \mathperiod
\end{equation}
The appearance of \emph{equal-time} commutation relations between fundamental operators is the key property of the deparametrised approach to a canonical quantisation of GFT, as proposed for a toy model in \cite{AGW18} and also suggested from a slightly different perspective on GFT in \cite{KO18}.
This is opposed to an algebra in which operators at different times $\chi^0$ and ${\chi^0}'$ commute, which had been assumed in previous work starting from \cite{GOS14}.

It is then desirable to express the field operators in terms of creation and annihilation operators that satisfy the commutation relations of the Heisenberg algebra and are Hermitian conjugates.
The time-dependent operators\footnote{Recall that for real fields we have $\op{\varphi}_J(\chi^0, -\vec{k})=\op{\varphi}^\dagger_J(\chi^0, \vec{k})$ and $\op{\pi}_J(\chi^0, -\vec{k})=\op{\pi}^\dagger_J(\chi^0, \vec{k})$.}
\begin{subequations}
  \label{eq:fock_operators}
  \begin{align}
    &
    \op{a}_J(\chi^0, \vec{k})
    =
    A_J(\vec{k})
    \op{\varphi}_J(\chi^0, \vec{k})
    +
    \frac{\imagi}{2 A_J(\vec{k})}
    \op{\pi}_J(\chi^0, \vec{k})
    \mathcomma
    \\
    &
    \hermconj{\op{a}}_J(\chi^0, \vec{k})
    =
    A_J(\vec{k})
    \op{\varphi}_J(\chi^0, - \vec{k})
    -
    \frac{\imagi}{2 A_J(\vec{k})}
    \op{\pi}_J(\chi^0, - \vec{k})
    \mathcomma
  \end{align}
\end{subequations}
where $A_J(\vec{k})$ is an arbitrary real function, satisfy the ladder operator commutation relations
\begin{equation}
  \commutator{
    \op{a}_J(\chi^0, \vec{k})
  }{
    \hermconj{\op{a}}_{J'}(\chi^0, \vec{k}')
  }
  =
  \delta_{JJ'}
  (2\pi)^d
  \diracdelta{\vec{k} - \vec{k}'}
  \mathperiod
\end{equation}

Let us assume that $A_J(\vec{k}) = A_J(- \vec{k})$.
Inserting \eqref{eq:fock_operators} into \eqref{eq:hamiltonian} gives
\begin{equation}
  \begin{aligned}
    \op{H}_J(\vec{k})
    =
    \frac{\gftkcoeff{2}}{2}
    \Bigg(
      &
      \left(
        \frac{A_J(\vec{k})^2}{
          \abs{\gftkcoeff{2}}^2
        }
        +
        \frac{
          \omega_J(\vec{k})^2
        }{
          4 A(\vec{k})^2
        }
      \right)
      \left(
        \op{a}_J(\vec{k})
        \op{a}_J(-\vec{k})
        +
        \hermconj{\op{a}}_J(\vec{k})
        \hermconj{\op{a}}_J(-\vec{k})
      \right)
      \\
      &
      -
      \left(
        \frac{A_J(\vec{k})^2}{
          \abs{\gftkcoeff{2}}^2
        }
        -
        \frac{
          \omega_J(\vec{k})^2
        }{
          4 A(\vec{k})^2
        }
      \right)
      \left(
        \op{a}_J(- \vec{k})
        \hermconj{\op{a}}_J(- \vec{k})
        +
        \hermconj{\op{a}}_J(\vec{k})
        \op{a}_J(\vec{k})
      \right)
    \Bigg)
    \mathcomma
  \end{aligned}
\end{equation}
where the operators are now to be viewed as being given in the Schrödinger picture: the Hamiltonian $\op{H}_J(\vec{k})$ does not evolve in time and hence can be evaluated in either Heisenberg or Schr\"odinger picture.
Up to now $A_J(\vec{k})$ is any even real function.
For $\omega_J(\vec{k}) \neq 0$ there is a natural choice for this function, namely
\begin{equation}
  A_J(\vec{k})
  =
  \sqrt{
    \frac{
      \abs{\omega_J(\vec{k})}
      \abs{\gftkcoeff{2}}
    }{
      2
    }
  }
  \mathcomma
\end{equation}
which leads to a simple expression for the resulting Hamiltonian.
Depending on the sign of $\omega_J(\vec{k})^2$ one then obtains either a harmonic oscillator Hamiltonian ($\omega_J(\vec{k})^2 < 0$) or a squeezing Hamiltonian ($\omega_J(\vec{k})^2 > 0$).
The first type of Hamiltonian is of the form
\begin{equation}
  \label{eq:oscillator_hamiltonian}
  \op{H}_J(\vec{k})
  =
  -\sign(\gftkcoeff{2})
  \frac{|\omega_J(\vec{k})|}{2}
      \left(
        \op{a}_J(- \vec{k})
        \hermconj{\op{a}}_J(- \vec{k})
        +
        \hermconj{\op{a}}_J(\vec{k})
        \op{a}_J(\vec{k})
      \right)
  \mathperiod
\end{equation}
Such a Hamiltonian has a stable ground state and would lead to a `static' cosmology in its geometric interpretation.
For a generic GFT model one would expect that there are modes with such Hamiltonians.
Notice that since $\omega_J(\vec{k})^2 = \vec{k}^2 + m_J^2$ this case requires that $m_J^2 < 0$ and $\abs{\vec{k}}<\sqrt{-m_J^2}$.

The second case of a squeezing Hamiltonian leads to an exponentially growing instability; hence, if there are such unstable modes, they are eventually going to give the dominant contribution to the overall dynamics.
The modes for which $\omega_J(\vec{k})^2 > 0$ are hence the modes of main interest, and in the following we will restrict ourselves to this case.
Note that even if $m_J^2 < 0$, all modes with large enough $\abs{\vec{k}}$ fall into this category.
The squeezing Hamiltonian we are going to consider is therefore given by
\begin{equation}
  \label{eq:squeezing_hamiltonian}
  \op{H}_J(\vec{k})
  =
  \sign(\gftkcoeff{2})
  \frac{\omega_J(\vec{k})}{2}
  \left(
    \op{a}_J(\vec{k})
    \op{a}_J(-\vec{k})
    +
    \hermconj{\op{a}}_J(\vec{k})
    \hermconj{\op{a}}_J(-\vec{k})
  \right)
  \mathperiod
\end{equation}
This is the type of Hamiltonian that generates squeezing in quantum optics, in the sense that it produces a squeezed state out of, e.g., the Fock vacuum.
The fact that cosmological evolution in a universe filled with a massless scalar field can be described as squeezing was the key insight of \cite{AGW18}.

\subsection{Equations of motion}
\label{sec:hamiltonian_eom}
Returning to the Heisenberg picture, the creation and annihilation operators have to obey the Heisenberg equations of motion
\begin{subequations}
  \begin{align}
    &
    \pdv{}{\chi^0}
    \op{a}_J(\chi^0, \vec{k})
    =
    - \imagi
    \commutator{
      \op{a}_J(\chi^0, \vec{k})
    }{
      \op{H}
    }
    \mathcomma
    \\
    &
    \pdv{}{\chi^0}
    \hermconj{\op{a}}_J(\chi^0, \vec{k})
    =
    - \imagi
    \commutator{
      \hermconj{\op{a}}_J(\chi^0, \vec{k})
    }{
      \op{H}
    }
    \mathperiod
  \end{align}
\end{subequations}
We now assume that all relevant modes have dynamics given by a squeezing type Hamiltonian \eqref{eq:squeezing_hamiltonian}. In this case the solutions are given by
\begin{subequations}
  \label{eq:fock_operators_timedep}
  \begin{align}
    &
    \op{a}(\chi^0, \vec{k})
    =
    \op{a}(\vec{k})
    \cosh(\omega_J(\vec{k}) \chi^0)
    -
    \imagi
    \sign(\gftkcoeff{2})
    \hermconj{\op{a}}_J(- \vec{k})
    \sinh(\omega_J(\vec{k}) \chi^0)
    \mathcomma
    \\
    &
    \hermconj{\op{a}}(\chi^0, \vec{k})
    =
    \hermconj{\op{a}}(\vec{k})
    \cosh(\omega_J(\vec{k}) \chi^0)
    +
    \imagi
    \sign(\gftkcoeff{2})
    \op{a}_J(- \vec{k})
    \sinh(\omega_J(\vec{k}) \chi^0)
    \mathperiod
  \end{align}
\end{subequations}
The time-dependent group field and conjugate momentum are therefore given by
\begin{subequations}
  \label{eq:gft_field_timedep}
  \begin{align}
    &
    \op{\varphi}_J(\chi^0, \vec{k})
    =
    \op{\varphi}_J(\vec{k})
    \cosh(\omega_J(\vec{k}) \chi^0)
    -
    \frac{1}{
      \omega_J(\vec{k})
      \gftkcoeff{2}
    }
    \op{\pi}_J(- \vec{k})
    \sinh(\omega_J(\vec{k}) \chi^0)
    \mathcomma
    \\
    &
    \op{\pi}_J(\chi^0, \vec{k})
    =
    \op{\pi}_J(\vec{k})
    \cosh(\omega_J(\vec{k}) \chi^0)
    -
    \omega_J(\vec{k})
    \gftkcoeff{2}
    \op{\varphi}_J(- \vec{k})
    \sinh(\omega_J(\vec{k}) \chi^0)
    \mathperiod
  \end{align}
\end{subequations}
In these equations, operators without explicit $\chi^0$ dependence denote operators at initial time $\chi^0=0$ or equivalently the operators defined in the Schr\"odinger picture.
Note that \eqref{eq:gft_field_timedep} is consistent with \eqref{eq:gft_field_conjugate_momentum}.
The exponential instability in the Hamiltonian dynamics is evident from these expressions.
An observable built from $\op{a}(\chi^0, \vec{k})$ and $\hermconj{\op{a}}(\chi^0, \vec{k})$ will typically diverge as $\expe^{N|\omega_J(\vec{k})\chi^0|}$ towards the past and future, where $N$ is the number of ladder operators appearing in the observable.
We will see explicit examples of this below.

\subsection{Symmetries and conserved quantities}
\label{sec:symmetries}
In classical field theory, Noether's theorem provides a relation between continuous symmetries of a system and its conserved quantities \cite{Noe18}.
In the following we write the action as
$
  S[\varphi]
  =
  \int
  \intmeasure[D]{\chi}
  \mathcal{L}(\varphi, \partial_\chi \varphi)
$
and refer to the scalar density $\mathcal{L}$ as the Lagrangian.
If the Lagrangian is invariant under the infinitesimal transformations
\begin{equation}
  \chi^\alpha \mapsto \chi^\alpha + \delta\chi^\alpha
  \mathcomma\quad
  \varphi_J \mapsto \varphi_J + \delta\varphi_J
  \mathcomma
\end{equation}
there is a divergenceless current
\begin{equation}
  J^\alpha
  =
  \sum_J
  \pdv{\mathcal{L}}{\left(\partial_\alpha \varphi_J\right)}
  \delta\varphi_J
  -
  \left(
    \sum_J
    \pdv{\mathcal{L}}{\left(\partial_\alpha \varphi_J\right)}
    \partial_\beta \varphi_J
    -
    \delta^\alpha_\beta
    \mathcal{L}
  \right)
  \delta\chi^\beta
  \mathcomma
\end{equation}
where $\partial_\alpha = \partial / \partial \chi_\alpha$ and the Einstein summation convention is used for the indices denoting the scalar field components.
The current gives rise to quantities that are conserved with respect to relational time
\begin{equation}
  Q
  =
  \int
  \intmeasure[d]{\vec{\chi}}
  J^0(\chi^\alpha)
  \mathcomma\quad
  \frac{\partial}{\partial \chi^0}
  Q
  =
  0
  \mathperiod
\end{equation}
Conserved quantities are crucial in the physical interpretation of a quantum state both in standard quantum field theory and in GFT.

On the quantum level we adopt a normal ordering prescription, such that all the annihilation operators are to be placed to the right of any creation operators.
Normal ordering is denoted by $\normalorder{\blank}$.
Notice that in the Heisenberg picture normal ordering has to be performed at the level of the time-dependent operators (as opposed to the operators evaluated at the initial time).

The simplest symmetries are translations $\chi^\alpha \mapsto \chi^\alpha + \epsilon^\alpha$. The Lagrangian \eqref{eq:kinetic_term_truncated} is invariant under translations and has a corresponding current $J^\alpha = \Theta^{\alpha\beta} \epsilon_\beta$ with
\begin{equation}
  \Theta_{\alpha\beta}
  =
  -
  \sum_J
  \pdv{\mathcal{L}}{\left(\partial^\alpha \varphi_J\right)}
  \partial_\beta \varphi_J
  +
  \delta_{\alpha\beta} \mathcal{L}
  =
  \sum_J
  \mathcal{K}^{(2)}_J
  \pdv{\varphi_J}{\chi^\alpha}
  \pdv{\varphi_J}{\chi^\beta}
  +
  \delta_{\alpha \beta} \mathcal{L}
  \mathperiod
\end{equation}
From this one can define the total momentum of the group field
\begin{equation}
  \op{P}_\alpha
  =
  \int
  \intmeasure[d]{\vec{\chi}}
  \normalorder{
    \op*{\Theta_{0\alpha}(\chi^\alpha)}
  }
  \mathperiod
\end{equation}
The $\op{P}_0$ component of this expression gives the total Hamiltonian (with an overall change in sign)
\begin{equation}
  \label{eq:total_energy}
  \op{P}_0
  =
  -
  \sum_J
  \sign(\gftkcoeff{2})
  \int
  \fourierintmeasure[d]{\vec{k}}
  \frac{\omega_J(\vec{k})}{2}
  \left(
    \op{a}_J(\vec{k})
    \op{a}_J(-\vec{k})
    +
    \hermconj{\op{a}}_J(\vec{k})
    \hermconj{\op{a}}_J(-\vec{k})
  \right)
\end{equation}
and the remaining `spatial momentum' components are given by
\begin{equation}
  \label{eq:total_momentum}
  \op{\vec{P}}
  =
  \sum_J
  \op{\vec{P}}_J
  =
  -
  \sum_J
  \int
  \intmeasure[d]{\vec{\chi}}
  \normalorder{
    \op{\pi}_J(\chi^\alpha)
    \vecgrad \op{\varphi}_J(\chi^\alpha)
  }
  =
  \sum_J
  \int
  \fourierintmeasure[d]{\vec{k}}
  \vec{k}\,
  \hermconj{\op{a}}_J(\vec{k})
  \op{a}_J(\vec{k})
  \mathperiod
\end{equation}
Writing the Lagrangian as $\mathcal{L}=\sum_J \mathcal{L}_J$ each $\mathcal{L}_J$ in the sum is separately invariant under arbitrary translations.
It follows that in the sums over $J$ appearing in the total Hamiltonian \eqref{eq:total_energy} and momentum \eqref{eq:total_momentum}, each $J$ component is already conserved individually.

While $\op{P}_0$ and $\op{\vec{P}}$ initially appear to be part of a $D$-dimensional vector, once they are expressed in terms of creation and annihilation operators we again observe the symmetry breaking exhibited by our Hamiltonian formalism: the components of $\op{\vec{P}}$ transform covariantly under rotations of the $d$ `spatial' scalar fields but momentum and Hamiltonian take quite different forms.
We will see below that the operators associated to `boosts', rotations that mix the `time' and `space' components of $\field{R}^D$, also take an unusual form in our setting and that the full set of conserved quantities still satisfies the algebra of the full Euclidean group $\liegroup{E}(D)$.
Notice also that the total momentum involves an integral over the number density $\hermconj{\op{a}}_J(\vec{k}) \op{a}_J(\vec{k})$ weighted by the wavenumber $\vec{k}$.
While neither the number density nor its integral over all $\vec{k}$ are conserved, time evolution by squeezing always creates pairs of particles with opposite wavenumber leaving the total momentum conserved.

To give a physical interpretation to the conserved quantities $\op{H}$ and $\op{\vec{P}}$, recall how in \cref{sec:kinetic_term} we motivated the form of the GFT kinetic term using symmetry arguments:
we pointed out that the Lagrangian for a free massless scalar field $\chi$ on an arbitrary spacetime background $g_{\mu\nu}$ is invariant under a shift symmetry $\chi \mapsto \chi+a$.
Of course according to Noether's theorem this shift symmetry is also associated with a conserved quantity, namely
\begin{equation}
  \int
  \intmeasure[d]{\vec{x}}
  \pi_\chi(x)
  =
  -\int
  \intmeasure[d]{\vec{x}}
  \sqrt{-g}
  \,
  g^{0\nu}
  \partial_\nu\chi
  \mathcomma
\end{equation}
the total canonical momentum of the scalar field $\chi$.

It is then very natural to interpret the conserved energy $\op{H}$ and momentum $\op{\vec{P}}$ in the GFT as corresponding to the conserved canonical momenta $(\pi_{\chi})_0$ and $\vec{\pi}_{\chi}$ associated to the $D$ scalar matter fields, and we will use this interpretation in what follows.
In the case where a GFT state can be interpreted as a spatially homogeneous geometry, this interpretation was already used in previous work such as \cite{AGW18,GP19} where the Hamiltonian was interpreted as the conjugate momentum of a single scalar matter field.
Note however that these conserved quantities exist both in the GFT formulated here and in usual scalar field theory on a fixed background regardless of whether the geometry is spatially homogeneous.

The Lagrangian \eqref{eq:kinetic_term_truncated} is also invariant under infinitesimal rotations $\chi^\alpha \mapsto \chi^\alpha + \delta\omega\indices{^\alpha_\beta} \chi^\beta$, where $\delta\omega^{\alpha\beta}$ is antisymmetric.
The associated current is $J^\alpha = \frac{1}{2} M^{\alpha\beta\gamma} \delta\omega_{\beta\gamma}$ with
\begin{equation}
  M_{\alpha\beta\gamma}
  =
  \Theta_{\alpha\beta} \chi_\gamma
  -
  \Theta_{\alpha\gamma} \chi_\beta
  \mathperiod
\end{equation}
The conserved quantity is an antisymmetric tensor
\begin{equation}
  \op{M}_{\alpha\beta}
  =
  \int
  \intmeasure[d]{\vec{\chi}}
  \normalorder{
    \op{M}_{0\alpha\beta}
  }
  =
  \int
  \intmeasure[d]{\vec{\chi}}
  (
    \normalorder{
      \op{
        \Theta
      }_{0\alpha}
      \chi_\beta
      -
      \op{
        \Theta
      }_{0\beta}
      \chi_\alpha
    }
  )
  \mathperiod
\end{equation}
Its $0i$-components are given by
\begin{equation}
  \op{M}_{0i}
  =
  -
  \frac{\imagi
  }{
    2
  }
  \sum_J
  \sign(\gftkcoeff{2})
  \int
  \fourierintmeasure[d]{\vec{k}}
  \omega_J(\vec{k})
  \left(
    \op{a}_J(- \vec{k})
    \pdv{}{k^i}
    \op{a}_J(\vec{k})
    -
    \hermconj{\op{a}}_J(- \vec{k})
    \pdv{}{k^i}
    \hermconj{\op{a}}_J(\vec{k})
  \right)
  -
  \op{P}_i
  \,
  \chi_0
  \mathcomma
\end{equation}
where $\op{P}_i$ is the $i$-component of the momentum given in \eqref{eq:total_momentum}.

The $ij$-components are reminiscent of the definition of an angular momentum and are given by
\begin{equation}
  \op{M}_{ij}
  =
  \imagi
  \sum_J
  \int
  \fourierintmeasure[d]{\vec{k}}
  \hermconj{\op{a}}_J(\vec{k})
  \left(
    k_i
    \pdv{}{k^j}
    -
    k_j
    \pdv{}{k^i}
  \right)
  \op{a}_J(\vec{k})
  \mathperiod
\end{equation}
By the same argument as given above for the case of translational symmetries, each $J$ component of these rotational charges is conserved separately.

We can again give a physical interpretation to these quantities by relating them to conserved quantities of a theory with $D$ massless scalar fields in a curved spacetime.
Here the analogous symmetry is under global infinitesimal `rotations' $\chi^\alpha \mapsto \chi^\alpha + \delta\omega\indices{^\alpha_\beta} \chi^\beta$ that mix these scalar fields, which leads to conserved quantities
\begin{equation}
  -\int
  \intmeasure[d]{\vec{x}}
  \sqrt{-g}
  \,
  g^{0\nu}
  \left(
  \chi^\beta
  \partial_\nu\chi^\alpha
  -
  \chi^\alpha
  \partial_\nu\chi^\beta
  \right)
  =
  \int
  \intmeasure[d]{\vec{x}}
  \left(
  \pi_\chi^\alpha(x)
  \chi^\beta(x)
  -
  \pi_\chi^\beta(x)
  \chi^\alpha(x)
  \right)
\end{equation}
for each pair $(\chi^\alpha,\chi^\beta)$ of scalar fields.
We will not make use of these `angular momentum' quantities in the following but they contain physical information about GFT states beyond that captured by the energy and momentum.

Finally, let us remark that the operators defined above satisfy the expected commutation relations of infinitesimal generators of the Euclidean group $\liegroup{E}(D)$, i.e.,
\begin{align}
  \commutator{
    \op{M}_{\alpha \beta}
  }{
    \op{M}_{\gamma \delta}
  }
  &
  =
  - \imagi
  \left(
    \delta_{\alpha\gamma}
    \op{M}_{\beta\delta}
    -
    \delta_{\alpha\delta}
    \op{M}_{\beta\gamma}
    -
    \delta_{\beta\gamma}
    \op{M}_{\alpha\delta}
    +
    \delta_{\beta\delta}
    \op{M}_{\alpha\gamma}
  \right)
  \mathcomma
  \\
  \commutator{
    \op{M}_{\alpha\beta}
  }{
    \op{P}_{\gamma}
  }
  &
  =
  - \imagi
  \left(
    \delta_{\alpha\gamma}
    \op{P}_{\beta}
    -
    \delta_{\beta\gamma}
    \op{P}_{\alpha}
  \right)
  \mathperiod
\end{align}
This shows that the Euclidean symmetry of the Lagrangian \eqref{eq:kinetic_term_truncated} is unbroken at the kinematical level in the Hamiltonian theory obtained from canonical quantisation.
We will however see that the use of certain states of interest in physical applications, such as coherent states, breaks the symmetry explicitly.
From the perspective of conserved quantities and symmetry generators this is because `spatial' rotations leave this class of states invariant whereas `boosts' do not, similar to how we introduced a preferred foliation for field configurations before quantisation.
Since coherent states are peaked on a classical phase-space configuration, it is not surprising that they inherit the broken symmetry that we have observed already in the classical theory.

\subsection{Observables}
The conserved quantities we have identified provide important observables that can be used to characterise GFT states.
Other observables with nontrivial time dependence can also be defined.
An important quantity is the total particle number
\begin{equation}
  \op{N}(\chi^0)
  =
  \sum_J
  \int
  \fourierintmeasure[d]{\vec{k}}
    \hermconj{\op{a}}_J(\chi^0, \vec{k})
    \op{a}_J(\chi^0, \vec{k})
  \mathperiod
\end{equation}
This is in general not conserved due to the particle creation by squeezing, so this observable has genuine dependence on relational time $\chi^0$ and we must use the time-dependent ladder operators when working in the Heisenberg picture.
The integrand $\hermconj{\op{a}}_J(\chi^0, \vec{k})    \op{a}_J(\chi^0, \vec{k})$ gives a particle number density associated to a `species' $J$ and wavenumber $\vec{k}$.

A geometric interpretation can be given to states in the GFT Fock space by importing operators from loop quantum gravity \cite{Ori13}.
Here the most basic object is the volume operator
\begin{equation}
  \label{eq:volume_operator}
  \op{V}(\chi^0)
  =
  \sum_J
  v_J
  \int
  \fourierintmeasure[d]{\vec{k}}
    \hermconj{\op{a}}_J(\chi^0, \vec{k})
    \op{a}_J(\chi^0, \vec{k})
\end{equation}
where $v_J$ is the volume eigenvalue associated to a single `quantum', i.e., a geometric tetrahedron or four-valent spin network vertex in LQG, with Peter--Weyl representation labels $J$ (see, e.g., \cite{BH11} for background).
This volume likewise has a nontrivial time dependence, which we will study below in the cosmological interpretation of our Hamiltonian dynamics.

Deparametrisation using the scalar field $\chi^0$ as a relational clock has the usual consequence that there can be no `time' operator associated to $\chi^0$: instead, all other observables are functions of $\chi^0$.
Our quantum theory has been defined so that all its observables are related to the possible readings of the clock $\chi^0$.

We could attempt to define operators corresponding to the values of the scalar fields $\vec\chi$.
Such operators would allow extracting relational information about the evolution of the $d$ other matter fields relative to $\chi^0$.
It should be clear that such operators would be the analogue of position operators in standard scalar field theory and come with the usual conceptual and practical difficulties associated with this concept.
We could use the analogue of Newton--Wigner position operators \cite{NW49} to define matter fields for a single tetrahedron.
The usual objection that Newton--Wigner operators do not transform well under boosts would not necessarily be an issue in our setting which is anyway only covariant with respect to rotations of the $d$ `spatial' fields.
We would then have to deal with other issues such as the rather unclear physical meaning of the `total scalar field value' given by the sum of all values for a scalar field in a many-particle configuration (see \cite{GO14,MO20} for related discussions).
We will not study these operators further in this paper, but they may be of interest in future work.

\subsection{Coherent states}
\label{sec:coherent_states}
When computing expectation values, a choice of initial state must be made.
A particularly interesting class of states are the (Fock) coherent states which can be defined as being the eigenstates of the annihilation operator at the initial time when Heisenberg and Schr\"odinger pictures agree.
These states have semiclassical behaviour and can be constructed so that relative uncertainties in energy and volume become very small at late times, which can be interpreted as the emergence of a classical universe in GFT \cite{GP19}.

In the formalism developed in this section coherent states can be defined as
\begin{equation}
  \ket{\sigma}
  =
  \expe^{- \norm{\sigma}^2 / 2}
  \exp
  \left(
    \sum_J
    \int
    \fourierintmeasure[d]{\vec{k}}
    \sigma_J(\vec{k})
    \hermconj{\op{a}}_J(\vec{k})
  \right)
  \ket{0}
  \mathcomma
\end{equation}
where $\ket{0}$ is the Fock vacuum and the norm $\norm{\cdot}$ is given by
\begin{equation}
  \norm{\sigma}^2
  =
  \sum_J
  \int
  \fourierintmeasure[d]{\vec{k}}
  \abs{\sigma_J(\vec{k})}^2
  \mathperiod
\end{equation}
Such coherent states are entirely characterised by complex functions $\sigma_J(\vec{k})$.
To be more specific, one often makes the assumption that only one or a few $J$ modes contribute, in which case the sum over $J$ would simplify accordingly.
Coherent states which are peaked at certain values of $\vec{k}$ can be modelled by functions
\begin{equation}
  \sigma_J(\vec{k})
  =
  \sum_i
  f_\epsilon(\vec{k} - \vec{k}_i)
  \tau(\vec{k}_i)
  \mathcomma
\end{equation}
where $f_\epsilon(\vec{k})$ depends on a parameter $\epsilon$ and is chosen such that
\begin{equation}
  \label{eq:coherent_envelope}
  \lim_{
    \epsilon \rightarrow 0
  }
  \int
  \fourierintmeasure[d]{\vec{k}}
  \abs{
    f_\epsilon(\vec{k})
  }^2
  \phi(\vec{k})
  =
  \phi(\vec{0})
\end{equation}
for a test function $\phi(\vec{k})$. An example of such a function is given by
\begin{equation}
  \label{eq:gauss_epsilon}
  f_\epsilon(\vec{k})
  =
  \left(
    \frac{4 \pi}{\epsilon^2}
  \right)^{d/4}
  \expe^{
    -
    \frac{ \vec{k}^2 }{ 2 \epsilon^2 }
  }
  \mathperiod
\end{equation}
For this particular choice of $f_\epsilon(\vec{k})$ it is possible to perform an expansion in powers of $\epsilon$
\begin{equation}
  \int
  \fourierintmeasure[d]{\vec{k}}
  \abs{
    f_\epsilon(\vec{k})
  }^2
  \phi(\vec{k})
  =
  \phi(\vec{0})
  +
  \frac{\epsilon^2}{4}
  (\vecgrad^2 \phi)(\vec{0})
  +
  \bigO{\epsilon^4}
  \mathperiod
\end{equation}

As can be seen from the solution to the equation of motion in \cref{sec:hamiltonian_eom}, the $(J, \vec{k})$-mode of an observable typically has a time dependence proportional to $\expe^{N |\omega_J(\vec{k}) \chi^0|}$ for integer $N$.
Since for any given $\vec{k}$ there will always be a $\vec{k}'$ for which $\omega_J(\vec{k}') > \omega_J(\vec{k})$, the contribution of the mode $(J, \vec{k}')$ to the observable will grow faster than that of  $(J, \vec{k})$.
Therefore, unless one chooses particularly fine-tuned initial conditions in which most modes are set exactly to zero at all times, the approximation in which only some distinguished modes $(J_i, \vec{k}_i)$ contribute is valid only for a certain finite time.
As a consequence of this one can also show that a state initially peaked at some value $\vec{k}_0$ will not remain peaked at that value at later times; rather the wave number $\vec{k}_{*}$ where the state is peaked is a function of time, $\vec{k}_* = \vec{k}_*(\chi^0)$.
For an observable that has a time dependence given by $\expe^{N \omega_J(\vec{k}) \chi^0}$ for a mode $(J, \vec{k})$, the time dependence of the peak can be obtained by solving
\begin{equation}
  \evalat*{\odv{}{\vec{k}}}_{\vec{k} = \vec{k}_*}
  \left(
    \abs{f_\epsilon(\vec{k} - \vec{k}_0)}^2
    \expe^{N \omega_J(\vec{k}) \chi^0}
  \right)
  =
  0
  \mathcomma
\end{equation}
where one can choose for definiteness $f_\epsilon$ to be given by \eqref{eq:gauss_epsilon}. Indeed, for that choice one finds in the limit $\vec{k}_0^2 \gg m_J^2$ that at time $\chi^0$ the peak lies at
\begin{equation}
  \vec{k}_*
  =
  \vec{k}_0
  \left(
    1 + \frac{N \epsilon^2 \chi^0}{2|\vec{k}_0|}
  \right)
  \mathperiod
\end{equation}

The above discussion shows that care has to be taken when discussing wave packets that have a finite width.
In the examples where sharply peaked coherent states are used in the following, we will content ourselves with the idealisation in which the state remains peaked at the same value of $\vec{k}_0$ for all times; equivalently, we follow the time evolution only up to a maximal time $\chi^0_{{\rm max}}\ll \frac{2|\vec{k}_0|}{N\epsilon^2}$.
By tuning $\epsilon$ this maximal time can be made as large as desired but it cannot be made infinite.
This fact does not pose additional difficulties beyond those discussed in \cref{sec:lagrangian_eom}: we already assumed that interactions in the GFT can be neglected and this assumption can only hold for a finite time, again due to the exponential instability in our theory.
One then needs to be careful and work with coherent states that are sufficiently sharply peaked so that the time dependence of the peak does not become significant within the regime of the nearly free theory.
   \section{Effective cosmology}
\label{sec:eff_cosmology}

We will now proceed with the main application of the canonical formulation of GFT, namely the extraction of an effective cosmological dynamics, similarly to what has been done in numerous examples before \cite{GOS14,OSW16,CPS16,PS17,AGW18,GP19,AK20}.
The main new element in the derivation we give in this paper is the use of a deparametrised form of GFT dynamics in a theory describing quantum gravity coupled to multiple scalar matter fields.
We will encounter various subtleties arising in such a derivation.

As in previous work \cite{GP19} our aim will be to derive general effective equations, interpreted as GFT Friedmann equations, that make some simplifying assumptions about the quantum state but are otherwise as general as possible.
Generally speaking, these equations involve the total spatial volume, given as an expectation value of the volume operator \eqref{eq:volume_operator}, and its derivative with respect to relational time given by the scalar field $\chi^0$.
Such equations, in the most general case, depend on an infinite number of initial condition parameters and as such will not admit a straightforward solution or interpretation in terms of an emergent effective cosmology, whose initial data are only given in terms of a few parameters.
One then restricts to more specific initial conditions: in particular, one can choose coherent states which have semiclassical behaviour.
The most drastic restriction usually made is to assume that only a single Peter--Weyl mode $J_0$ of the expansion \eqref{eq:pw_decomp} is relevant for the effective dynamics.
This assumption can be partially motivated by a dynamical argument \cite{Gie16} saying that in a wide class of models a single mode will dominate asymptotically, but it is generally a somewhat \emph{ad hoc} choice made for computational convenience.
States that only excite a single mode $J_0$ describe a configuration of geometric quanta of identical `shape', which one might see as the analogue of spatial homogeneity in a discrete setting such as GFT.
In hindsight, this interpretation is corroborated by the agreement of effective GFT dynamics with classical Friedmann equations for homogeneous geometries that was found for such states in \cite{GP19} and many other examples.
In the present framework in which multiple matter fields are present, states and operators additionally depend on wavenumbers $\vec{k}$ associated to these extra degrees of freedom.
We will then similarly assume that the chosen state is peaked on a single wavenumber $\vec{k}_0$, or pair of wavenumbers $\pm\vec{k}_0$ (the second choice being in line with the dynamics of squeezing which generates pairs of particles with $\pm\vec{k}$).
Such states are then candidates for spatially homogeneous geometries in this extended setting.

Rather than restricting to a single $J_0$, in \cite{OSW16} `isotropic' GFT states associated to a single representation $j$ were constructed out of modes that have the form
\begin{equation}
  \varphi_j=\sum_{\multindex{m},\iota}\compconj{\alpha}_j^\iota \compconj{\mathcal{I}}^{\multindex{j}, \iota}_{\multindex{m}}\varphi_{\multindex{j}, \multindex{m}, \iota}
\end{equation}
where we have reinstated the explicit form $J = (\multindex{j}, \multindex{m}, \iota)$ of the multi-index $J$.
On the right-hand side one chooses the `isotropic form' $\multindex{j}=(j,j,j,j)$ and thus the left-hand side is specified by a single spin $j$; the weights $\alpha_j^\iota$ are chosen to make the resulting state into an eigenstate of the LQG volume operator with maximal eigenvalue.
We will not use this isotropic form in the following but our general results could be adapted to this choice.

\subsection{Regularised volume operator}
\label{sec:volume_regularisation}
The basic idea behind the cosmological application of GFT is to identify the volume of the Universe with an expectation value of the GFT volume operator and therefore with a weighted sum over the number operators which count the number of excited GFT quanta of a specific type.
Explicitly, we define the volume operator as
\begin{equation}
  \label{eq:volume_op}
  \op{V}(\chi^0)
  =
  \sum_J
  \int
  \fourierintmeasure[d]{\vec{k}}
  \op{V}_J(\chi^0, \vec{k})
  \mathcomma
\end{equation}
where the `partial volume densities' are defined as
\begin{equation}
  \op{V}_J(\chi^0, \vec{k})
  =
  v_J
  \,
  \hermconj{\op{a}}_J(\chi^0, \vec{k})
  \op{a}_J(\chi^0, \vec{k})
  \mathperiod
\end{equation}
This of course reproduces the definition for the GFT volume operator given in \eqref{eq:volume_operator}.
Since in our truncation to quadratic GFT dynamics all $J$ and all $\vec{k}$ modes are decoupled, one can study the dynamics of each $\op{V}_J(\chi^0, \vec{k})$ separately.

Using \eqref{eq:fock_operators_timedep} one finds
\begin{equation}
  \label{eq:volume_timedep}
  \begin{aligned}
    \op{V}_J(\chi^0, \vec{k})
    =
    &
    \frac{1}{2}
    \left(
      \op{V}_J(\vec{k}) - \op{V}_J(- \vec{k}) - v_J c_\infty
    \right)
    +
    \frac{1}{2}
    \left(
      \op{V}_J(\vec{k}) + \op{V}_J(- \vec{k}) + v_J c_\infty
    \right)
    \cosh(2 \omega_J(\vec{k}) \chi^0)
    \\
    &
    +
    \sign(\gftkcoeff{2})
    \frac{\imagi}{2}
    v_J
    \left(
      \op{a}_J(\vec{k})
      \op{a}_J(- \vec{k})
      -
      \hermconj{\op{a}}_J(\vec{k})
      \hermconj{\op{a}}_J(- \vec{k})
    \right)
    \sinh(2 \omega_J(\vec{k}) \chi^0)
    \mathcomma
  \end{aligned}
\end{equation}
where $\op{V}_J(\vec{k}) = v_J\hermconj{\op{a}}_J(\vec{k}) \op{a}_J(\vec{k})$ is the partial volume density operator at initial time and $c_\infty$ is a formally divergent quantity as explained in more detail below.

The total volume takes the somewhat simpler form
\begin{equation}
\label{eq:total_volume}
  \begin{aligned}
    \op{V}(\chi^0)
    =
    \sum_J
    \int
    \fourierintmeasure[d]{\vec{k}}
    \Big[
      -\frac{v_J c_\infty}{2}
      +
      \left(
        \op{V}_J(\vec{k}) + \frac{v_J c_\infty}{2}
      \right)
      &
      \cosh(2 \omega_J(\vec{k}) \chi^0)
      \\
      &
      \quad
      +
      \op{X}_J(\vec{k})
      \sinh(2 \omega_J(\vec{k}) \chi^0)
    \Big]
    \mathcomma
  \end{aligned}
\end{equation}
with the shorthand
\begin{equation}
  \label{eq:x_definition}
  \op{X}_J(\vec{k})
  =
  \sign(\gftkcoeff{2})
  \frac{\imagi}{2}
  v_J
  \left(
    \op{a}_J(\vec{k})
    \op{a}_J(- \vec{k})
    -
    \hermconj{\op{a}}_J(\vec{k})
    \hermconj{\op{a}}_J(- \vec{k})
  \right)
  \mathperiod
\end{equation}
The quantity $c_\infty$ introduced in \eqref{eq:volume_timedep} is equal to the contact term
\begin{equation}
    c_\infty
    =
    [\op{a}_J(\vec{k}),\hermconj{\op{a}}_J(\vec{k})]
\end{equation}
which formally diverges, $c_\infty=(2\pi)^d \evalat{\diracdelta{\vec{k}}}_{\vec{k} = \vec{0}}$.
The total volume, and in fact already each partial volume density, would appear to be infinite for any time other than $\chi^0=0$.
The volume operator hence requires regularisation in order to describe a meaningful quantity.

Physically, the divergence in $\op{V}_J(\chi^0, \vec{k})$ arises because we have now defined a squeezed state in a quantum field theory, which has an infinite number of degrees of freedom (even for each individual representation label $J$).
Each of the field modes gets squeezed by the Hamiltonian \eqref{eq:squeezing_hamiltonian}.
The analogue of such a state in (continuum) quantum optics describes a stationary light beam which is likewise associated with infinite particle number \cite{BLPS90}.

Standard regularisation prescriptions in quantum field theory amount to setting $c_\infty \rightarrow 0$.
For instance, one could replace \eqref{eq:volume_op} by a regularised volume operator
\begin{equation}
  \op{V}_{\mathsubscript{reg}}(\chi^0)
  =
  \lim_{\epsilon\rightarrow 0}
  \sum_J
  v_J
  \left(
  \int
  \fourierintmeasure[d]{\vec{k}}
  \fourierintmeasure[d]{\vec{k}'}
  \chi_\epsilon(\vec{k}-\vec{k}')
  \hermconj{\op{a}}_J(\chi^0, \vec{k})
  \op{a}_J(\chi^0, \vec{k}')
  -
  \text{divergent terms}
  \right)
  \mathcomma
\end{equation}
where $\chi_\epsilon(\vec{k}-\vec{k}')$ is the regularisation of a delta distribution such as a Gaussian, which depends on a parameter $\epsilon$ (see the discussion in \cref{sec:coherent_states}), and we subtract terms that diverge as $\epsilon\rightarrow 0$.
Since $\lim_{\epsilon\rightarrow 0}\chi_\epsilon(0)=\infty$ we need to set $c_\infty\rightarrow 0$.
Such a regularised volume operator only counts particles relative to the `background' which is already generated by squeezing the vacuum.
Indeed, an alternative regularisation prescription (which leads to the same result) would be to subtract the vacuum contribution, i.e.,
\begin{equation}
  \op{V}_{\mathsubscript{reg}}(\chi^0)
  =
  \op{V}(\chi^0)
  -
  \bra{0}
  \op{V}(\chi^0)
  \ket{0}
  \mathcomma
\end{equation}
where $\ket{0}$ is the Fock vacuum.
The proposal is then that only this regularised quantity is relevant for describing an effective dynamics for the emergent geometry.

We will leave $c_\infty$ general in the following in order to study the impact of this regularisation procedure on the effective cosmological dynamics.
Notice that for $c_\infty\neq 0$ the total volume \eqref{eq:total_volume} contains a divergent term
\begin{equation}
    \sum_J
      \frac{v_J c_\infty}{2}
    \int
    \fourierintmeasure[d]{\vec{k}}
      \left(
      \cosh(2 \omega_J(\vec{k}) \chi^0)
      -
      1
      \right)
\end{equation}
which requires additional regularisation, e.g., by cutoffs in the $\vec{k}$ integral and sum over $J$.

\subsection{Effective Friedmann equation}
\label{sec:eff_friedmann}
In the case of a classical flat Friedmann--Lema\^{i}tre--Robertson--Walker (FLRW) metric coupled to $D = d + 1$ massless scalar fields $\chi^0, \chi^1, \dotsc, \chi^d$ the Friedmann equation can be written as
\begin{equation}
  \label{eq:friedmann_flrw}
  \left(
    \frac{
      V'(\chi^0)
    }{
      V(\chi^0)
    }
  \right)^2
  =
  12 \pi G
  \left(
    1
    +
    \frac{
      (\pi_\chi)_{1}^2
      +
      \dotsb
      +
      (\pi_\chi)_{d}^2
    }{
      (\pi_\chi)_{0}^2
    }
  \right)
  \mathcomma
\end{equation}
where the volume $V$ is expressed as a function of the massless scalar field $\chi^0$, $(\pi_\chi)_\alpha$ is the momentum of the scalar field $\chi^\alpha$ and $G$ is Newton's constant.
To see explicitly how to obtain \eqref{eq:friedmann_flrw} from the usual form of the Friedmann equation, one can first write the latter in terms of a volume variable $V(t)\propto a^3(t)$ where $a(t)$ is the usual scale factor and $t$ is a general time variable (not necessarily proper time).
In this notation the usual general relativistic (first) Friedmann equation is
\begin{equation}
  \label{eq:general_friedmann}
  \left(
    \frac{
      V'(t)
    }{
      N(t)V(t)
    }
  \right)^2
  =
  24 \pi G
  \rho(t)
  =
  12 \pi G
  \left(
    \frac{
      (\pi_\chi)_{0}^2
      +
      (\pi_\chi)_{1}^2
      +
      \dotsb
      +
      (\pi_\chi)_{d}^2
    }{
      V(t)^2
    }
  \right)
  \mathcomma
\end{equation}
where the total matter energy density is $\rho(t)=\sum_\alpha\rho_\alpha(t)$ and $\rho_\alpha(t) = (\pi_\chi)_\alpha^2/(2V(t)^2)$ for each of the $D$ massless scalar fields.
To fix the lapse function $N(t)$, specify to the case where $t\equiv\chi^0$ is identified with the value of one of the scalar fields.
Using the definition of the conjugate momentum
\begin{equation}
  (\pi_\chi)_0
  =
  \frac{1}{N(t)}
  \odv{\chi^0(t)}{t}
  V(t)
  \mathcomma
\end{equation}
we see that if $t\equiv\chi^0$ we have $N(t)=V(t)/(\pi_\chi)_0$.
Inserting this relation into \eqref{eq:general_friedmann} yields  \eqref{eq:friedmann_flrw}.
In the previously studied case \cite{OSW16} where only a single matter field was present, the right-hand side of \eqref{eq:friedmann_flrw} was simply a constant but here it depends on the values of all conserved momenta $(\pi_\chi)_\alpha$.

An effective Friedmann equation for GFT is obtained by determining the analogous expression for expectation values of the GFT volume operator \eqref{eq:volume_op}.
Here, by differentiating \eqref{eq:total_volume} with respect to $\chi^0$ and using \eqref{eq:volume_timedep} before taking expectation values, we find that
\begin{equation}
  \label{eq:effective_friedmann}
  \begin{aligned}
    \expval{
      &
      \op{V}'(\chi^0)
    }^2
    =
    \sum_{J,J'}
    \int
    \fourierintmeasure[d]{\vec{k}}
    \fourierintmeasure[d]{\vec{k}'}
    4
    \omega_J(\vec{k})
    \omega_{J'}(\vec{k}')
    \\
    &
    \times
    \sqrt{
      \left(
        \expval{
          \op{V}_{J}(\chi^0, \vec{k})
        }
        -
        \expval{
          \op{V}_{J}(\vec{k})
        }
      \right)
      \left(
        \expval{
          \op{V}_{J}(\chi^0, \vec{k})
        }
        +
        \expval{
          \op{V}_{J}(- \vec{k})
        }
        +
        v_J c_{\infty}
      \right)
      +
      \expval{
        \op{X}_{J}(\vec{k})
      }^2
    }
    \\
    &
    \times
    \sqrt{
      \left(
        \expval{
          \op{V}_{J'}(\chi^0, \vec{k}')
        }
        -
        \expval{
          \op{V}_{J'}(\vec{k}')
        }
      \right)
      \left(
        \expval{
          \op{V}_{J'}(\chi^0, \vec{k}')
        }
        +
        \expval{
          \op{V}_{J'}(- \vec{k}')
        }
        +
        v_J c_{\infty}
      \right)
      +
      \expval{
        \op{X}_{J'}(\vec{k}')
      }^2
    }
    \mathperiod
  \end{aligned}
\end{equation}
This effective Friedmann equation is completely general, but does not admit a straightforward cosmological interpretation as it depends on the initial condition parameters $\op{V}_{J}(\vec{k})$ and $\op{X}_{J}(\vec{k})$ in a rather complicated way; taking the quotient $    \expval{\op{V}'(\chi^0)}^2 / \expval{\op{V}(\chi^0)}^2$ will not lead to an expression that can be simplified easily.

\subsection{Late-time limit}
\label{sec:late_times}
The first consistency check for the effective cosmological dynamics in our model is whether they reduce to those of general relativity at large volume or low energy density.

The time-dependent expression \eqref{eq:volume_timedep} shows that each expectation value $\expval{\op{V}_{J}(\chi^0, \vec{k})}$ grows exponentially for early or late times if the mode is excited at all initially (the only other possibility is to set $c_\infty=0$ and choose the Fock vacuum with $\expval{\op{V}_{J}(\vec{k})}=\expval{\op{X}_{J}(\vec{k})}=0$, in which case the mode does not contribute to the dynamics).
This implies that for each relevant mode, at a sufficiently early or late time
\begin{equation}
  \expval{\op{V}_{J}(\chi^0, \vec{k})}
  \gg
  \expval{\op{V}_{J}(\pm\vec{k})}
  \mathcomma
  \quad
  \expval{\op{V}_{J}(\chi^0, \vec{k})}
  \gg
  \expval{\op{X}_{J}(\vec{k})}
  \mathperiod
\end{equation}

In this limit any dependence of the effective Friedmann equation \eqref{eq:effective_friedmann} on these initial condition parameters then disappears, and one obtains in the limit of large $\abs{\chi^0}$
\begin{equation}
\label{eq:asymptotic_friedmann}
  \left(
    \frac{
      \expval{
        \op{V}'(\pm\infty)
      }
    }{
      \expval{
        \op{V}(\pm\infty)
      }
    }
  \right)^2
  =
  \frac{
    \sum_{J,J'}
    \int
    \fourierintmeasure[d]{\vec{k}}
    \fourierintmeasure[d]{\vec{k}'}
    4
    \omega_J(\vec{k})
    \omega_{J'}(\vec{k}')
    \expval{
      \op{V}_J(\pm\infty, \vec{k})
    }
    \expval{
      \op{V}_{J'}(\pm\infty, \vec{k}')
    }
  }{
    \sum_{J,J'}
    \int
    \fourierintmeasure[d]{\vec{k}}
    \fourierintmeasure[d]{\vec{k}'}
    \expval{
      \op{V}_J(\pm\infty, \vec{k})
    }
    \expval{
      \op{V}_{J'}(\pm\infty, \vec{k}')
    }
  }
  \mathperiod
\end{equation}
The limit of infinite $\chi^0$ which is suggested by the notation is an idealisation since, as we already stressed a few times, our quadratic Hamiltonian dynamics will break down at some finite time when GFT interactions will become relevant.
Nevertheless, for states whose dynamics are described by the free GFT dynamics for a sufficiently long period of time, this asymptotic form will be a good approximation once all relevant modes have been sufficiently excited by squeezing.

We can simplify the asymptotic form \eqref{eq:asymptotic_friedmann} of the effective dynamics further in the special case for which only one or two modes $(J_0, \pm \vec{k}_0)$ are relevant, the case in which these states could be interpreted as (approximately) spatially homogeneous geometries.
In this case, the function $\omega_J(\vec{k})$ can be approximated by its value $\omega_{J_0}(\vec{k}_0)$ for these dominant modes, and taken out of the integral and sum.
\Eqref{eq:asymptotic_friedmann} then simplifies to
\begin{equation}
  \label{eq:simple_latetime}
  \left(
    \frac{
      \expval{
        \op{V}'(\pm\infty)
      }
    }{
      \expval{
        \op{V}(\pm\infty)
      }
    }
  \right)^2
  \approx
  4
  \omega_{J_0}(\vec{k}_0)^2
  =
  4
  (m_{J_0}^2 + \vec{k}_0^2)
  \mathperiod
\end{equation}
One would now like to verify that these asymptotic dynamics are compatible with the dynamics of an FLRW universe in general relativity, as was done in previous work for GFT models involving a single scalar matter field.
First, in order to reproduce this previously studied case, we can consider the limit in which the contribution from the scalar fields $\chi^1,\ldots,\chi^d$ is negligible.
This amounts to assuming $\vec{k}_0 \ll m_{J_0}^2$ in \eqref{eq:simple_latetime}.
The dynamics of general relativity can be recovered in this case if one makes the identification $m_{J_0}^2 = 3\pi G$ of the GFT coupling parameter $m_{J_0}$ with the classical Newton's constant $G$, as was done in previous work  \cite{OSW16,OSW17,Gie16,AGW18,GP19}.

In the more general case, the right-hand side of \eqref{eq:simple_latetime} depends on the quantity $\vec{k}_0$ which also needs to be given a cosmological interpretation.
The expectation value of the GFT momentum \eqref{eq:total_momentum} is approximately
\begin{equation}
   \label{eq:approx_momentum}
  \expval{
  \op{\vec{P}}
  }
  =
  \int
  \fourierintmeasure[d]{\vec{k}}
  \vec{k}
  \expval{
  \hermconj{\op{a}}_{J_0}(\vec{k})
  \op{a}_{J_0}(\vec{k})
  }
  \approx
  \vec{k}_0
  \left(
      \expval{
        \op{N}_+
      }
      -
      \expval{
        \op{N}_-
      }
  \right)
  \mathcomma
\end{equation}
where $\op{N}_+$ and $\op{N}_-$ denote the initial number of quanta in the mode $+\vec{k}_0$ and $-\vec{k}_0$, respectively.
(Notice that the operators appearing in this equation are the operators at time $\chi^0=0$.)
As discussed in \cref{sec:symmetries} the GFT momentum $\op{\vec{P}}$ can be interpreted as corresponding to the conjugate momentum $\vec{\pi}_\chi$ of the $d$ `spatial' matter fields, which appears in the classical Friedmann equation \eqref{eq:friedmann_flrw}.
Likewise, the energy $\op{H}$ is interpreted as the conjugate momentum of the `clock' field $\chi^0$.

We can then write \eqref{eq:simple_latetime} as
\begin{equation}
   \label{eq:simple_latetime2}
  \left(
    \frac{
      \expval{
        \op{V}'(\pm\infty)
      }
    }{
      \expval{
        \op{V}(\pm\infty)
      }
    }
  \right)^2
  \approx
  4m_{J_0}^2
  \left(
  1 + \frac{\expval{
     \op{\vec{P}}
      }^2}{
  \left(
      \expval{
        \op{N}_+
      }
      -
      \expval{
        \op{N}_-
      }
  \right)^2
      m_{J_0}^2}
  \right)
  \mathperiod
\end{equation}
The combination $\expval{\op{N}_+}-\expval{\op{N}_-}$ is related to the choice of initial conditions, but it does not have a direct interpretation in terms of (cosmological) late-time dynamics.
For some more specific states, $(\expval{\op{N}_+}-\expval{\op{N}_-})^2 m_{J_0}^2$ is approximately equal to the energy squared
\begin{equation}
      \expval{
        \op{H}
      }^2
  \approx
  \omega^2_{J_0}(\vec{k}_0)
  \left(
  \int
  \fourierintmeasure[d]{\vec{k}}
  \frac{1}{2}
  \left\langle
    \op{a}_{J_0}(\vec{k})
    \op{a}_{J_0}(-\vec{k})
    +
    \hermconj{\op{a}}_{J_0}(\vec{k})
    \hermconj{\op{a}}_{J_0}(-\vec{k})
  \right\rangle
  \right)^2
\end{equation}
but this is not true in general, as we will see shortly when specifying to coherent states.
In the general case, one can then only bring \eqref{eq:simple_latetime2} into the suggestive form
\begin{equation}
   \label{eq:suggestive_latetime}
   \left(
    \frac{
      \expval{
        \op{V}'(\pm\infty)
      }
    }{
      \expval{
        \op{V}(\pm\infty)
      }
    }
  \right)^2
  \approx
  4m_{J_0}^2
  \left(
  1 + \kappa^2 \frac{\expval{
     \op{\vec{P}}
      }^2}
      {\expval{
        \op{H}
      }^2}
  \right)
\end{equation}
where $\kappa$ is a function of initial conditions and can, but does not need to be, close to one.
Agreement with the classical Friedmann dynamics \eqref{eq:friedmann_flrw} at late times or large volume is then only reached for rather specific initial states.
The underlying reason for this perhaps puzzling result is again the symmetry breaking of our Hamiltonian formalism from $\liegroup{E}(D)$ to the smaller group $\liegroup{E}(d)\times\liegroup{E}(1)$: while in the approximation \eqref{eq:approx_momentum} the momentum $\expval{\op{\vec{P}}}$ is directly related to the initial number of particles in different modes, the same is not true for the energy which is an \emph{a priori} independent initial condition parameter.

There is a possible, though at this stage rather tentative, physical interpretation of the appearance of $\kappa\neq 1$ in \eqref{eq:suggestive_latetime} as including the effect of inhomogeneities in the geometry.
We focused on states peaked on only one or two modes $(J_0, \pm \vec{k}_0)$ which are the best candidates for describing spatially homogeneous geometries.
However, interpreting the scalars $\vec{\chi}$ as relational spatial coordinates as suggested in \cite{GOW18,Gie18} would seem to imply that states with $\vec{k}_0\neq \vec{0}$ should rather be thought of as describing `periodically inhomogeneous' effective geometries, whose inhomogeneities take the form of a plane wave.
(In the somewhat related setting of \cite{GO18} only $\vec{k}_0 = \vec{0}$ was considered as a homogeneous background for perturbations.)
Similarly, scalar fields $\vec\chi$ that can serve as useful relational coordinates cannot be exactly homogeneous in space.
Exploring this idea further would rely on a clearer understanding of inhomogeneities in the setting of GFT coupled to multiple matter scalar fields, which we must leave to future work.
In \cref{app:periodic_inhomo} we show an example in classical general relativity for which periodic inhomogeneities can modify the coupling of the matter energy density appearing in the effective Friedmann equation, but the relation of this example to GFT cosmology is not yet clear.

\subsection{Single and double mode coherent states}
Up to now the discussion was independent of any specific choice of initial state.
We will now discuss several special cases for such an initial state and consider the general form of the effective dynamics, including also nontrivial quantum geometry effects.
In the following we will only consider the regularised volume operator with $c_\infty = 0$, that is we neglect all the contributions from the `vacuum' geometry, in order to avoid further regularisation and divergence issues (cf.\ the discussion in \cref{sec:volume_regularisation}).

The time-dependent expression \eqref{eq:total_volume} for the total volume simplifies drastically if we assume again that the initial state is sharply peaked on two modes $(J_0,\pm\vec{k}_0)$:
\begin{equation}
\label{eq:simple_volume}
  \begin{aligned}
    \expval{
      \op{V}(\chi^0)
    }
    &
    =
    \sum_J
    \int
    \fourierintmeasure[d]{\vec{k}}
    \Big[
        \expval{
          \op{V}_J(\vec{k})
        }
      \cosh(2 \omega_J(\vec{k}) \chi^0)
      +
      \expval{
        \op{X}_J(\vec{k})
      }
      \sinh(2 \omega_J(\vec{k}) \chi^0)
    \Big]
  \\
    &
    \approx
      \expval{
        \op{V}
      }
      \cosh(2 \omega_{J_0}(\vec{k}_0) \chi^0)
      +
      \expval{
        \op{X}
      }
      \sinh(2 \omega_{J_0}(\vec{k}_0) \chi^0)
    \mathcomma
  \end{aligned}
\end{equation}
where $\op{V} = \sum_J \int \fourierintmeasure[d]{\vec{k}} \op{V}_J(\vec{k})$ is the total volume at initial time and we defined
\begin{equation}
  \op{X}
  =
  \sum_J
  \int
  \fourierintmeasure[d]{\vec{k}}
  \op{X}_J(\vec{k})
  =
  \sum_J
  \sign(\gftkcoeff{2})
  \frac{\imagi\,v_J}{2}
  \int
  \fourierintmeasure[d]{\vec{k}}
  \left(
    \op{a}_J(\vec{k})
    \op{a}_J(- \vec{k})
    -
    \hermconj{\op{a}}_J(\vec{k})
    \hermconj{\op{a}}_J(- \vec{k})
  \right)
  \mathperiod
\end{equation}

The simplified form \eqref{eq:simple_volume} satisfies a straightforwardly derived Friedmann equation, which corresponds to \eqref{eq:effective_friedmann} in this specific case.

First consider the even more specific case of a `single mode' state for which only one mode $(J_0, \vec{k}_0)$ is relevant.
In this case we find that
\begin{equation}
  \label{eq:effective_friedmann_singlemode}
  \frac{
  \expval[\singlemode]{
    \op{V}'(\chi^0)
  }^2
  }{
    \expval[\singlemode]{\op{V}(\chi^0)}^2
  }
  =
  4 \omega_{J_0}(\vec{k}_0)^2
  \left(
    1
    -
    \frac{
    \expval[\singlemode]{\op{V}}^2
    }{
     \expval[\singlemode]{\op{V}(\chi^0)}^2
    }
  \right)
  \mathperiod
\end{equation}
Notice that for this type of state the expectation values of both $\op{X}$ and $\op{H}$ vanish; thus, higher curvature corrections of the type seen in loop quantum cosmology, which would give a term proportional to the matter energy density inside the brackets in \eqref{eq:effective_friedmann_singlemode}, cannot be seen in this approximation in which the energy density is zero.
Such a single mode state can be realised by a coherent state with a coherent state function given by
\begin{equation}
  (\sigma_\singlemode)_J(\vec{k})
  =
  \delta_{J J_0}
  f_\epsilon(\vec{k} - \vec{k}_0)
  \tau(\vec{k}_0)
  \mathcomma
\end{equation}
where $f_\epsilon(\vec{k})$ is defined in \eqref{eq:coherent_envelope} and it is to be understood that all corrections involving powers of $\epsilon$ are neglected at the end.
The single mode coherent state functions are characterised by the value of the complex quantity $\tau(\vec{k}_0)$.
In terms of this quantity the expectation value of $\op{N}$ is given by
\begin{equation}
  \expval[\singlemode]{
    \op{N}
  }
  =
  \abs{\tau(\vec{k}_0)}^2
\end{equation}
and the initial volume is $\expval[\singlemode]{\op{V}}=v_{J_0}   \expval[\singlemode]{\op{N}}$.

Now  consider a more general `double mode' state where two modes $(J_0, \pm \vec{k}_0)$ are relevant.
For this type of state the effective Friedmann equation is given by
\begin{equation}
\label{eq:eff_friedmann_dmstate}
  \frac{
  \expval[\doublemode]{
    \op{V}'(\chi^0)
  }^2}{
    \expval[\doublemode]{\op{V}(\chi^0)}^2
  }
  =
  4 \omega_{J_0}(\vec{k}_0)^2
  \left(
    1
    -
    \frac{
    \expval[\doublemode]{\op{V}}^2
    }{
    \expval[\doublemode]{\op{V}(\chi^0)}^2
    }
    +
    \frac{
    \expval[\doublemode]{\op{X}}^2
    }{
    \expval[\doublemode]{\op{V}(\chi^0)}^2
    }
  \right)
  \mathperiod
\end{equation}
Such a double mode state can be realised by a coherent state with coherent state function
\begin{equation}
  (\sigma_\doublemode)_J(\vec{k})
  =
  \delta_{J J_0}
  (
    f_\epsilon(\vec{k} - \vec{k}_0)
    \tau(\vec{k}_0)
    +
    f_\epsilon(\vec{k} + \vec{k}_0)
    \tau(- \vec{k}_0)
  )
  \mathperiod
\end{equation}
The double mode coherent state function is characterised by two complex parameters, $\tau(\pm \vec{k}_0)$.
In terms of these parameters the expectation values of the operators $\op{N}$ and $\op{X}$ are given by (again $\expval[\doublemode]{\op{V}}=v_{J_0}   \expval[\doublemode]{\op{N}}$)
\begin{align}
  &
  \expval[\doublemode]{
    \op{N}
  }
  =
  \abs{
    \tau(\vec{k}_0)
  }^2
  +
  \abs{
    \tau(- \vec{k}_0)
  }^2
  \mathcomma
  \\
  &
  \expval[\doublemode]{
    \op{X}
  }
  =
  \imagi
  v_{J_0}
  \sign(\gftkcoeff[J_0]{2})
  \left(
    \tau(\vec{k}_0)
    \tau(-\vec{k}_0)
    -
    \compconj{\tau}(\vec{k}_0)
    \compconj{\tau}(-\vec{k}_0)
  \right)
  \mathperiod
\end{align}
The values of the relevant components of the Hamiltonian and the total momentum for such a double mode coherent state are
\begin{align}
  &
  \expval[\doublemode]{
    \op{H}
  }
  =
  \sign(\gftkcoeff[J_0]{2})
  \omega_{J_0}(\vec{k}_0)
  \left(
    \tau(\vec{k}_0)
    \tau(-\vec{k}_0)
    +
    \compconj{\tau}(\vec{k}_0)
    \compconj{\tau}(-\vec{k}_0)
  \right)
  \mathcomma
  \\
  &
  \expval[\doublemode]{
    \op{\vec{P}}
  }
  =
  \vec{k}_0
  \left(
    \abs{
      \tau(\vec{k}_0)
    }^2
    -
    \abs{
      \tau(- \vec{k}_0)
    }^2
  \right)
  \mathperiod
\end{align}

For these double mode coherent states, we can now compare the combination $(\expval{\op{N}_+}-\expval{\op{N}_-})^2 m_{J_0}^2$ with the energy squared $\expval{\op{H}}^2$ as discussed at the end of \cref{sec:late_times}.
Separating the coherent state parameters $\tau(\pm \vec{k}_0)$ into modulus and phase as suggested in \cite{OSW16}, we have $\tau(\pm \vec{k}_0) = \rho_\pm \expe^{\imagi\theta_\pm}$ and
\begin{align}
  \expval[\doublemode]{
    \op{H}
  }^2
  &
  =
  4
  \left(
  m_{J_0}^2
  +
  \vec{k}_0^2
  \right)
  \left(
    \rho_{+} \rho_{-}
  \right)^2
  \cos^2(\theta_{+} + \theta_{-})
  \mathcomma
  \\
  \left(
  \expval[\doublemode]{
    \op{N}_+
  }
  -
  \expval[\doublemode]{
    \op{N}_-
  }
  \right)^2
  m_{J_0}^2
  &
  =
  m_{J_0}^2
  \left(
    \rho_{+}^2
    -
    \rho_{-}^2
  \right)^2
  \mathperiod
\end{align}
Clearly one can find states with $  \expval[\doublemode]{ \op{H} }^2 \approx   \left(\expval[\doublemode]{\op{N}_+} - \expval[\doublemode]{\op{N}_-}\right)^2 m_{J_0}^2$, which is required for a late-time limit consistent with general relativity, but this is not the generic situation.
Such states exist only if the moduli $\rho_+$ and $\rho_-$ of the coherent state parameters are comparable.
For instance, in the limit $\vec{k}_0^2\ll m_{J_0}^2$ the late-time limit is consistent with general relativity for states that satisfy
\begin{equation}
  \abs{
    \cos(\theta_{+} + \theta_{-})
  }
  =
  \frac{
    \abs{
      \rho_{+}^2
      -
      \rho_{-}^2
    }
  }{
    2\rho_{+} \rho_{-}
  }
\end{equation}
which implies the necessary condition
\begin{equation}
  (\sqrt{2}-1)\rho_{+}
  \le
  \rho_{-}
  \le
  (\sqrt{2}+1)\rho_{+}
  \mathperiod
\end{equation}

For double mode coherent states we can bring \eqref{eq:eff_friedmann_dmstate} into a different form with a more direct cosmological interpretation. Namely, for such states
\begin{equation}
  \frac{
  v_{J_0}^2}{
  \omega_{J_0}(\vec{k}_0)^2
  }
  \expval[\doublemode]{
    \op{H}
  }^2
  +
  \frac{
  v_{J_0}^2}{
  \vec{k}_0^2
  }
  \expval[\doublemode]{
    \op{\vec{P}}
  }^2
  =
  \expval[\doublemode]{
    \op{V}
  }^2
  -
  \expval[\doublemode]{
    \op{X}
  }^2
\end{equation}
and \eqref{eq:eff_friedmann_dmstate} can also be written as
\begin{equation}
\label{eq:eff_friedmann_clearer}
  \frac{
  \expval[\doublemode]{
    \op{V}'(\chi^0)
  }^2}{
    \expval[\doublemode]{\op{V}(\chi^0)}^2
  }
  =
  4 \omega_{J_0}(\vec{k}_0)^2
  \left(
    1
    -
    \frac{
    v_{J_0}^2
    \expval[\doublemode]{
      \op{\vec{P}}
    }^2
    }{
    \vec{k}_0^2
    \expval[\doublemode]{\op{V}(\chi^0)}^2
    }
    -
    \frac{
    v_{J_0}^2
    \expval[\doublemode]{\op{H}}^2
    }{
    \omega_{J_0}(\vec{k}_0)^2
    \expval[\doublemode]{\op{V}(\chi^0)}^2
    }
  \right)
  \mathperiod
\end{equation}
Using the definitions
\begin{equation}
  \rho_{\chi^0}(\chi^0)
  =
  \frac{
     \expval[\doublemode]{\op{H}}^2
   }{
     2
     \expval[\doublemode]{\op{V}(\chi^0)}^2
   }
  \mathcomma
  \quad
  \rho_{\vec{\chi}}(\chi^0)
  =
  \frac{
    \expval[\doublemode]{\op{\vec{P}}}^2
  }{
    2
    \expval[\doublemode]{\op{V}(\chi^0)}^2
  }
\end{equation}
of the respective energy densities of the `clock' scalar field $\chi^0$ and of the remaining matter scalar fields (cf.~the discussion of a classical Friedmann equation in \eqref{eq:general_friedmann}), we can then write this effective GFT Friedmann equation in the form
\begin{equation}
\label{eq:friedmann_energy_dens}
  \frac{
  \expval[\doublemode]{
    \op{V}'(\chi^0)
  }^2}{
    \expval[\doublemode]{\op{V}(\chi^0)}^2
  }
  =
  4 \omega_{J_0}(\vec{k}_0)^2
  \left(
    1
    -
    \frac{
    2v_{J_0}^2
    }{
    \vec{k}_0^2
    }
    \rho_{\vec{\chi}}(\chi^0)
    -
    \frac{
    2v_{J_0}^2
    }{
    \omega_{J_0}(\vec{k}_0)^2
    }
    \rho_{\chi^0}(\chi^0)
  \right)
  \mathperiod
\end{equation}
This form can now be compared with effective Friedmann equations derived in previous work on GFT cosmology or, e.g., with cosmological equations that incorporate LQG corrections \cite{Tav08}.
In particular, we can write \eqref{eq:friedmann_energy_dens} more suggestively as
\begin{equation}
  \frac{
  \expval[\doublemode]{
    \op{V}'(\chi^0)
  }^2}{
    \expval[\doublemode]{\op{V}(\chi^0)}^2
  }
  =
  4 \omega_{J_0}(\vec{k}_0)^2
  \left(
    1
    -
    \frac{
      \rho_{\vec{\chi}}(\chi^0)
    }{
      \rho_{\vec{\chi},\critical} (\vec{k}_0)
    }
    -
    \frac{
      \rho_{\chi^0}(\chi^0)
    }{
      \rho_{\chi^0,\critical} (\vec{k}_0)
    }
  \right)
\end{equation}
where we defined $\rho_{\vec{\chi},\critical} (\vec{k}_0) = \frac{\vec{k}_0^2}{2v_{J_0}^2}$ and $\rho_{\chi^0,\critical} (\vec{k}_0) = \frac{\omega_{J_0}(\vec{k}_0)^2}{2v_{J_0}^2}$ as `critical densities' for the scalar fields $\chi^0$ and $\vec\chi$.
Correction terms of the form $-\rho(\chi^0)/\rho_\critical$ for some critical (Planckian) energy density $\rho_\critical$ are also found in effective equations for loop quantum cosmology, where they characterise the leading order deviations from general relativity \cite{Tav08,AS11}, so it is very interesting that they are of the same form in this GFT setting.
The critical densities depend on the wavenumber $\vec{k}_0$, in addition to their dependence on the representation $J_0$ that was already found in previous studies with only a single matter field \cite{OSW16,AGW18,GP19}.
The critical density for the clock field $\chi^0$ reduces in the limit $\vec{k}_0^2 \ll m_{J_0}^2$ to the expression
\begin{equation}
  \evalat*{
    \rho_{\chi^0,\critical} (\vec{k}_0)
  }_{\vec{k}_0\rightarrow 0}
  =
  \frac{
    m_{J_0}^2
  }{
    2v_{J_0}^2
  }
  =
  \frac{
  3\pi}{
  2}
  \rho_{\planck}
  \left(
    \frac{v_{\planck}
    }{
    v_{J_0}
    }
  \right)^2
\end{equation}
found in these previous works on GFT cosmology, where $\rho_{\rm P}$ is the Planck energy density and $v_{\rm P}$ the Planck volume.
Modes for which $\vec{k}_0^2$ is not negligible have a higher critical density.
For the matter fields $\vec{\chi}$ the critical density has no lower bound and can take any positive value.

We then see that the Universe undergoes a bounce when $\frac{\rho_{\vec{\chi}}(\chi^0)}{\rho_{\vec{\chi},\critical}(\vec{k}_0)}  + \frac{\rho_{\chi^0}(\chi^0)}{\rho_{\chi^0,\critical}(\vec{k}_0)}  = 1$ and that it cannot reach the classical singularity for generic states, i.e., $\expval[\doublemode]{\op{V}(\chi^0)}>0$ for all times unless $ \rho_{\vec{\chi}} = \rho_{\chi^0} = 0$.
Moreover, the different critical densities for the clock field and for the `spatial' matter fields $\vec\chi$ again show that our formalism breaks the classical symmetry under exchange of these matter fields.
If we work with a fixed pair of modes $(J_0, \pm \vec{k}_0)$, these critical densities take the same values for all coherent states for these modes.
Hence, for any such state one will find that the scalar field $\chi^0$ encounters a bounce at an energy density different from any of the other scalar fields.
This means that even for an initial coherent state chosen to reproduce the correct late-time physics, the impact of these high-energy corrections on phenomenology would be different for $\chi^0$ and the other scalar fields.
For high momentum modes with $\vec{k}_0^2\gg m_{J_0}^2$ the critical densities coincide for all fields.

An interesting difference between the effective Friedmann equation \eqref{eq:eff_friedmann_clearer} and analogous equations found previously in the deparametrised Hamiltonian setting for GFT \cite{AGW18,GP19} is that the latter contained a correction of the form $v_0/\expval{\op{V}(\chi^0)}$, whereas we only find corrections scaling as $\expval{\op{V}(\chi^0)}^{-2}$.
(A similar term scaling as $\expval{\op{V}(\chi^0)}^{-1}$ is also seen in mean-field derivations of effective dynamics \cite{OSW16,CPS16}.)
The reason for the absence of a $\expval{\op{V}(\chi^0)}^{-1}$ correction in our framework here can be traced back to the need for regularisation of the volume operator and our choice $c_\infty=0$.
This need for regularisation only occurs in a quantum field theory framework with an infinite number of degrees of freedom, whereas in these previous works there was only a single matter field used as clock and the formalism was analogous to standard quantum mechanics, where no regularisation is required.
In previous works in the deparametrised setting \cite{AGW18,GP19}, there was a constant contribution to the expression for the volume $\expval{\op{V}(\chi^0)}$ coming from a commutator $\commutator{\op{a}}{\hermconj{\op{a}}}$ which then also appeared in the effective Friedmann equations.
We have set this contribution to zero.
(In \cite{AGW18} it was suggested to remove this $v_0/\expval{\op{V}(\chi^0)}$ term by using a symmetrised definition of the volume operator, which has a similar effect.)

\subsection{Comparison with mean-field treatment for complex field}

Our results on effective cosmological dynamics are an interesting extension of known results on the effective cosmological dynamics of GFT models that include a single matter scalar field: for the first time, we have studied the impact of including multiple scalar fields into such dynamics.
We could see that at least for some rather specific choices of initial state (coherent states peaked on only a single representation label $J_0$ and a single pair $\pm\vec{k}_0$ of wavenumbers with further restrictions on initial data), the dynamics at large volume reduce to general relativity, while at high energies there are corrections to general relativity dynamics that lead to a bounce and singularity resolution for generic initial states.

The results were also puzzling: why do the clock field $\chi^0$ and the other fields $\vec\chi$ appear on a different footing in the effective dynamics, given that the initial GFT was perfectly symmetric under permutations or rotations of all matter fields?
We already argued in \cref{sec:lagrangian_eom,sec:distinguished_field} that this symmetry breaking arises from treating the elliptic equation of motion as an initial value problem, which introduces a preferred slicing of the scalar field configuration space $\field{R}^D$ such that initial data are regular (e.g., square-integrable) on each slice.
Such symmetry breaking should then not be specific to our formalism of deparametrised canonical quantisation, but be visible more generally.
To investigate this question further, we now compare our results with a mean-field analysis of the cosmological dynamics for GFT with a complex field, as used very successfully in \cite{OSW16,OSW17} and in most other existing work on GFT cosmology.
We will use the techniques developed in \cite{Gie16} for deriving general solutions and generalise some results given there.

Working in the same approximations as we have throughout this paper, we assume that the Peter--Weyl components of the GFT field satisfy the same equation of motion \eqref{eq:eom}.
Working again in a decomposition into Fourier modes (and thus making again the assumption that fields at `constant time' admit such a decomposition, which singles out a preferred direction of `time'), we then have
\begin{equation}
  \left(
    \frac{\partial^2}{\partial(\chi^0)^2}
    -
    \vec{k}^2
    -
    m_J^2
  \right)
  \varphi_J(\chi^0,\vec{k})
  =
  0
\end{equation}
whose general solution can be given straightforwardly,
\begin{equation}
  \varphi_J(\chi^0,\vec{k})
  =
  \alpha_J(\vec{k})
  \expe^{\sqrt{\vec{k}^2+m_J^2}\chi^0}
  +
  \beta_J(\vec{k})
  \expe^{-\sqrt{\vec{k}^2+m_J^2}\chi^0}
\end{equation}
where $ \alpha_J(\vec{k}) $ and $\beta_J(\vec{k})$ specify the initial conditions.
Recall that only for this subsection we work with a complex GFT field, so $\alpha_J(\vec{k})$ and $\beta_J(\vec{k})$ are arbitrary complex numbers.

In the usual mean-field treatment one now defines the total volume as (see, e.g., \cite{GO18})
\begin{equation}
  \begin{aligned}
    V(\chi^0)
    &
    =
    \sum_J
    v_J
    \int
    \intmeasure[d]{\vec{\chi}}
    \abs{\varphi_J(\chi^0,\vec{\chi})}^2
    \\
    &
    =
    \begin{aligned}[t]
      \sum_J
      v_J
      \int
      \fourierintmeasure[d]{\vec{k}}
      \Big(
        &
        \abs{\alpha_J(\vec{k})}^2
        \expe^{2\sqrt{\vec{k}^2+m_J^2}\chi^0}
        \\
        &
        +
        2\Re\left[
          \compconj{\beta}_J(\vec{k})
          \alpha_J(\vec{k})
        \right]
        +
        \abs{\beta_J(\vec{k})}^2
        \expe^{-2\sqrt{\vec{k}^2+m_J^2}\chi^0}
      \Big)
    \end{aligned}
  \end{aligned}
\end{equation}
and, for states sharply peaked on a single $J_0$ and a pair of wavenumbers $\pm\vec{k}_0$, we obtain approximate late-time dynamics
\begin{equation}
   \left(
   \frac{
   V'(\chi^0)
   }{
   V(\chi^0)
   }
   \right)^2
   \approx
   4
   \left(
   m_{J_0}^2+\vec{k}_0^2
   \right)
   =
   4 m_{J_0}^2
   \left(
   1
   +
   \frac{\vec{k}_0^2
   }{
   m_{J_0}^2
   }
   \right)
\end{equation}
which  unsurprisingly agrees with what we had found in \eqref{eq:simple_latetime}.
As in \cref{sec:late_times}, considering the case $\vec{k}_0^2\ll m_{J_0}^2$ in which the contribution from the `spatial' matter fields is negligible enforces that $m_{J_0}^2 = 3 \pi G$ in order to have agreement with classical general relativity.
One would then like to rewrite the term $\frac{\vec{k}_0^2}{m_{J_0}^2}$ using conserved quantities with a cosmological interpretation, as we have done in our earlier analysis.

Following the proposal of \cite{OSW16}, we could define as conjugate momentum of the clock $\chi^0$
\begin{equation}
  \begin{aligned}
    (\pi_\chi)_{0}
    &
    =
    -
    \frac{\imagi}{2}
    \sum_J
    \int
    \intmeasure[d]{\vec{\chi}}
    \left(
      \compconj{\varphi}_J(\chi^0,\vec{\chi})
      \pdv{
        \varphi_J(\chi^0,\vec{\chi})
      }{
        \chi^0
      }
      -
      \varphi_J(\chi^0,\vec{\chi})
      \pdv{
        \compconj{\varphi}_J(\chi^0,\vec{\chi})
      }{
        \chi^0
      }
    \right)
    \\
    &
    =
    \sum_J
    \int
    \fourierintmeasure[d]{\vec{k}}
    2\sqrt{\vec{k}^2+m_J^2}
    \,
    \Im\left[
      \compconj{\beta}_J(\vec{k})
       \alpha_J(\vec{k})
    \right]
  \end{aligned}
\end{equation}
which is manifestly conserved under time evolution.
However, as detailed in \cite{OSW16}, the conservation of this quantity is not due to time-translation invariance, but due to a global $\liegroup{U}(1)$ symmetry of the GFT with a complex field.
In the present context where there are now multiple matter scalar fields, this leads to the issue that there is only a single conserved quantity associated to $\liegroup{U}(1)$ symmetry, but $D \ge 2$ matter fields.
The analogously defined
\begin{equation}
   \frac{\imagi}{2}
   \sum_J
   \int
   \intmeasure[d]{\vec{\chi}}
   \left(
     \compconj{\varphi}_J(\chi^0,\vec{\chi})
     \vecgrad\varphi_J(\chi^0,\vec{\chi})
     -
     \varphi_J(\chi^0,\vec{\chi})
     \vecgrad\compconj{\varphi}_J(\chi^0,\vec{\chi})
   \right)
\end{equation}
does \emph{not} satisfy a conservation law.
The physical interpretation of such a `conjugate momentum', which grows exponentially at large volume, would be rather unclear.

Repeating instead the discussion of \cref{sec:symmetries} and using Noether's theorem for a complex GFT field leads to the definition of a conserved momentum
\begin{equation}
  \begin{aligned}
   \vec{P}
   &
   =
   \frac{1}{2}
   \sum_J
   \gftkcoeff{2}
   \int
   \intmeasure[d]{\vec{\chi}}
   \left(
     \pdv{
       \compconj{\varphi}_J(\chi^0,\vec{\chi})
     }{
       \chi^0
     }
     \vecgrad\varphi_J(\chi^0,\vec{\chi})
     +
     \pdv{
       \varphi_J(\chi^0,\vec{\chi})
     }{
       \chi^0
     }
     \vecgrad\compconj{\varphi}_J(\chi^0,\vec{\chi})
   \right)
   \\
   &
   =
   \sum_J
   \gftkcoeff{2}
   \int
   \fourierintmeasure[d]{\vec{k}}
   2\vec{k}
   \,
   \sqrt{\vec{k}^2+m_J^2}
   \,
   \Im\left[
     \compconj{\beta}_J(\vec{k})
     \alpha_J(\vec{k})
   \right]
  \end{aligned}
\end{equation}
and a conserved energy
\begin{equation}
  \begin{aligned}
   E
   &
   =
   \frac{1}{2}
   \sum_J
   \gftkcoeff{2}
   \int
   \intmeasure[d]{\vec{\chi}}
   \left(
     -
     \abs*{
       \pdv{
         \varphi_J(\chi^0,\vec{\chi})
       }{
         \chi^0
       }
     }^2
     +
     \abs*{
       \pdv{
         \varphi_J(\chi^0,\vec{\chi})
       }{
         \vec\chi
       }
     }^2
     +
     m_J^2
     \abs*{
       \varphi_J(\chi^0,\vec{\chi})
     }^2
   \right)
   \\
   &
   =
   \sum_J
   \gftkcoeff{2}
   \int
   \fourierintmeasure[d]{\vec{k}}
   2
   \left(
     \vec{k}^2+m_J^2
   \right)
   \,
   \Re\left[
     \compconj{\beta}_J(\vec{k})
     \alpha_J(\vec{k})
   \right]
  \end{aligned}
\end{equation}
and we reach a similar conclusion as in \cref{sec:late_times}: even for states peaked on a single mode or pair of modes, the GFT energy and momentum are independent initial conditions.
In this case, this can be seen by the fact that they depend, respectively, on the imaginary and real parts of the combination $\compconj{\beta}_J(\vec{k})\alpha_J(\vec{k})$, which are \emph{a priori} unrelated.
For generic states, the ratio $\frac{\vec{k}_0^2}{m_{J_0}^2}$ appearing in the effective dynamics is then not necessarily close to the ratio $\frac{\vec{P}^2}{E^2}$ that one would expect in order to match with general relativity at late times.

The fact that the symmetry breaking observed throughout this paper occurs also in a mean-field treatment illustrates that it is not due to the use of a deparametrised approach to canonical quantisation, but is present already in the classical theory once one introduces an initial value problem for the elliptic differential equation \eqref{eq:eom}.

   \section{Discussion}
\label{sec:discussion}

In this paper we have derived a canonical quantisation of group field theory models for quantum gravity coupled to an arbitrary number $D\ge 2$ of free massless scalar fields, focusing on a deparametrised approach where one of the scalar fields is chosen to play the role of a clock before quantisation.
We saw how the resulting theory generalises the previously formulated deparametrised Hamiltonian GFT for a single scalar field \cite{WEw19,GPW19}.
Throughout the paper we restricted ourselves to a quadratic action, neglecting the impact of GFT interactions on the dynamics.
We defined creation and annihilation operators from the group field and its conjugate momentum which act on a GFT Fock space, and we found conserved quantities such as energy, momentum and angular momentum using Noether's theorem.
These conserved quantities are important in the physical interpretation of the Hamiltonian GFT since they can be identified with conserved quantities of a theory of $D$ massless scalar fields on an arbitrary curved spacetime.

We used symmetry arguments and a derivative expansion (first proposed in \cite{OSW16} for models with a single matter field) to restrict the quadratic GFT actions we studied: all $D$ matter fields should be minimally coupled to gravity, and appear in the action on the same footing.
This implied that our actions were symmetric under the Euclidean group $\liegroup{E}(D)$ of rotations and translations in all $D$ matter fields.
The resulting equations of motion then involve a Laplacian on $\field{R}^D$, and are hence of elliptic type.
Viewing our dynamics as evolution relative to one of the scalar fields $\chi^0$ means reformulating these dynamical equations as an evolution problem in one direction in $\field{R}^D$.
For an elliptic partial differential equation, this initial value problem is not well-posed: the resulting dynamics lead to instabilities by which small initial perturbations will grow exponentially.
These instabilities are desirable from the GFT perspective, since they allow the `growth' of an emergent geometry out of an initial state; in particular, they allow for a realistic cosmological dynamics for which an initially small universe filled with scalar matter can grow exponentially.
In the Hamiltonian setting, such dynamics lead to a squeezing Hamiltonian which, likewise, can create squeezed states with arbitrarily high numbers of quanta from the Fock vacuum \cite{AGW18,WEw19}.
This implies that our assumption of free dynamics is only an approximation that holds for specific states for a finite time before higher-order terms in the action become relevant (see, e.g., \cite{GP19}).
In particular, the limit of infinite times, in which the fields diverge, cannot be taken consistently.

A fundamental issue we then encountered is that such a choice of `time' direction leads to a breaking of the original symmetry under $\liegroup{E}(D)$ to a smaller group $\liegroup{E}(d)\times\liegroup{E}(1)$ containing rotations among $d=D-1$ fields but not rotations that mix these $d$ fields with the `clock field'.
From the perspective of the field configurations we consider, these are assumed to be regular (e.g., square-integrable) on each hyperplane $\field{R}^d$ orthogonal to the $\chi^0$ direction but are not regular on any other slicing of $\field{R}^D$.
This symmetry breaking, which can be understood already classically, has numerous consequences.
Most notably, we saw its impact on the effective cosmological dynamics for such a Hamiltonian GFT.
The simplest cosmological dynamics were derived assuming that the quantum state is sharply peaked on two out of the infinitely many field modes of the GFT, and that it can be described by a coherent state in these modes.
Under these approximations, we found that the cosmological dynamics of GFT can reduce to the cosmological dynamics of general relativity at late times, but that this requires further restrictions on initial conditions, unlike what was found previously for a single matter field \cite{OSW16,AGW18,WEw19,GP19}.
The reason is that the conserved quantities in GFT that represent conjugate momenta to the scalar fields take different forms for the `clock' field and for the other matter fields.
Only specific states satisfy the relations between these different quantities that are needed to obtain the correct late-time dynamics.
We also found that generic states lead to a resolution of the big bang singularity by a bounce at high energies, but that the phenomenology of the bounce distinguishes between the matter fields, leading to different `critical' energy densities at which the bounce can occur for the different fields.
We clarified that the difference in the properties of the `clock' matter field and the other matter fields is not related to our use of deparametrisation but can be seen at the mean-field level where only classical field configurations appear and no explicit deparametrisation is used.
Thus, we could extend existing results on the cosmological effective dynamics of GFT in an interesting but rather subtle way to theories that include multiple scalar matter fields.

The GFT dynamics we have formulated then  breaks the symmetry under arbitrary linear  $\liegroup{E}(D)$ transformations that map between different possible `clock' choices.
This is the first time that the issue of covariance under coordinate changes (here only global ones) was studied in a Hamiltonian setting for GFT, and our results should be of interest for the development of the GFT formalism.
Some simpler models in quantum cosmology are known to have inequivalent quantum corrections when different choices of time coordinate are compared \cite{GD84,Mal16,BH16}.
In GFT one does not directly quantise a theory of gravity, and the issue is harder to formalise; but it would be important to develop a more covariant setting for GFT in which one is free to change between different clock choices, e.g., inspired by the general setting proposed in \cite{Van18,Hoe18} and further expanded and generalised in \cite{HSL19,HSL20}.
These papers introduce advanced methods in order to map between different deparametrised theories and, e.g., to define a generalised notion of observable corresponding to `time'.
However, these papers deal with low-dimensional quantum systems and the extension to a field theory such as GFT seems far from straightforward.
Moreover, we saw that the symmetry breaking in our setting already happens when formulating the equations of motion as an initial value problem, and as such seems intrinsic to any canonical formulation, including of constrained (Dirac) type.
It is then not easy to see how a more covariant Hamiltonian formulation could ever arise in such models.

In general most work on GFT is done in the path integral approach \cite{Fre05,Ori06,Kra11,Ori12}, which does not \emph{a priori} require singling out a direction for time evolution.
The results in this paper suggest that the path integral might be more suitable for leading to effective GFT dynamics that treat all matter fields on equal footing.
A symmetry breaking analogous to what we have seen in the Hamiltonian setting could arise from a choice of boundary conditions in the path integral: if these are set on $d$-dimensional hyperplanes, the path integral would again single out a preferred (orthogonal) direction.
Alternatively, boundary conditions could be defined on the boundary of a region in $\field{R}^D$ that respects the symmetries of the Laplacian, such as a $D$-ball.
Such a formulation would also be in line with the nature of the equations of motion as elliptic partial differential equations.
It is less clear how such boundary conditions would be interpreted from the perspective of effective cosmology, where one usually thinks of homogeneous hypersurfaces evolving in time.

The results in this paper formulate some of the first steps into extending the GFT formalism to models which contain a large number of dynamical matter fields coupled to quantum gravity.
Understanding better the interplay of geometric and matter degrees of freedom will be crucial for bringing GFT models closer to reality, and clarifying the physical predictions they might make in cosmology and other observationally relevant settings.

  \begin{acknowledgments}
    We thank Edward Wilson-Ewing for comments on the first arXiv version of the manuscript.
    The work of SG was funded by the Royal Society through a University Research Fellowship (UF160622) and a Research Grant for Research Fellows (RGF\textbackslash R1\textbackslash 180030).
    AP was supported by the same Research Grant for Research Fellows awarded to SG.
  \end{acknowledgments}

  \appendix
  \section{Diagonal kinetic term in GFT}
\label{app:diagonal_kinetic}
In this appendix we discuss the diagonalisation of a common class of kinetic terms for the GFT action \eqref{eq:action_general} for both complex and real group fields.

Consider a complex group field
\begin{equation}
  \varphi:
  \liegroup{SU}(2)^4
  \times
  \field{R}
  \rightarrow
  \field{C}
  \mathcomma
  \quad
  (g_a, \chi)
  \mapsto
  \varphi(g_a, \chi)
  \mathperiod
\end{equation}
In \eqref{eq:pw_decomp} the Peter--Weyl type decomposition of the group field is given as
\begin{equation}
  \label{eq:pw_decomp_2}
  \varphi  (g_a, \chi^\alpha)
  =
  \sum_J
  \varphi_J (\chi^\alpha)
  D_J(g_a)
  \mathcomma
\end{equation}
with the multi-index $J = (\multindex{j}, \multindex{m}, \iota)$.
The class of kinetic terms we consider is given by
\begin{equation}
\label{eq:kinetic-class}
  \mathcal{K}[\varphi]
  =
  \sum_J
  \int
  \intmeasure{\chi}
  \intmeasure{\chi'}
  \compconj{\varphi}_J(\chi)
  \mathcal{K}_J(\chi, \chi')
  \varphi_J(\chi')
  \mathcomma
\end{equation}
where we consider only the case in which the kernel $\mathcal{K}(\chi, \chi')$ is symmetric, $\mathcal{K}(\chi, \chi') = \mathcal{K}(\chi', \chi)$.
This form is motivated by considering kinetic terms that, when written in the original $g_a$ variables, are local in these variables and only depend on their derivatives, as in \cite{OSW16,OSW17} for complex fields and in \cite{GPW19} for real fields.
Next, we split the components $\varphi_J$ into their (rescaled) real and imaginary parts
\begin{align}
  \varphi_J(\chi)
  =
  \frac{1}{\sqrt{2}}
  \left(
    \varphi_{1, J}(\chi)
    +
    \imagi
    \varphi_{2, J}(\chi)
  \right)
  \mathperiod
\end{align}
Note that a real group field $\varphi  (g_a, \chi)$ would not have $\varphi_{2, J} = 0$ since the Wigner $D$-matrices appearing in \eqref{eq:pw_decomp_2} are not real.
The kinetic term is then diagonal,
\begin{equation}
  \mathcal{K}[\varphi]
  =
  \frac{1}{2}
  \sum_J
  \int
  \intmeasure{\chi}
  \left(
    \varphi_{1, J}(\chi)
    \mathcal{K}_J(\chi, \chi')
    \varphi_{1, J}(\chi)
    +
    \varphi_{2, J}(\chi)
    \mathcal{K}_J(\chi, \chi')
    \varphi_{2, J}(\chi)
  \right)
  \mathcomma
\end{equation}
where the off-diagonal terms cancel each other by virtue of the symmetry of $\mathcal{K}(\chi, \chi')$.

For a real group field the Peter--Weyl modes must satisfy the reality condition \eqref{eq:group_field_reality_condition},
\begin{equation}
  \compconj{\varphi}_J(\chi)
  =
  \epsilon_J
  \varphi_{-J}(\chi)
  \mathcomma
\end{equation}
where $-J = (\multindex{j}, - \multindex{m}, \iota)$ and $\epsilon_J = (-1)^{\sum_i (j_i - m_i)}$.
This implies that in the case of a real group field the components for distinct $J$ are not independent, rather they must satisfy
\begin{equation}
  \varphi_{1, -J}(\chi)
  =
  \epsilon_J
  \varphi_{1, J}(\chi)
  \mathcomma
  \quad
  \varphi_{2, -J}(\chi)
  =
  -
  \epsilon_J
  \varphi_{2, J}(\chi)
  \mathperiod
\end{equation}

This shows that the kinetic term \eqref{eq:kinetic-class} can always be brought into the form
\begin{equation}
  \mathcal{K}[\varphi]
  =
  \frac{1}{2}
  \sum_J
  \int \intmeasure{\chi}
  \varphi_J(\chi)
  \mathcal{K}_J(\chi, \chi')
  \varphi_J(\chi')
\end{equation}
with suitable redefinitions of $J$ and $\mathcal{K}_J(\chi, \chi')$.
   \section{Universe with periodic inhomogeneity}
\label{app:periodic_inhomo}

In \cref{sec:eff_friedmann} we presented the Friedmann equation for a flat FLRW universe in classical general relativity coupled to a number of free massless scalars.
This classical equation was compared with what we obtained as effective cosmological dynamics in GFT, in particular we checked for agreement of effective GFT dynamics and classical cosmology at late times.

In this appendix we slightly generalise the discussion to include possible effects of inhomogeneities.
As we explained in the main text, this generalisation could be relevant in giving a physical interpretation to the GFT states we used in deriving effective cosmological dynamics, which were peaked on a wavenumber $\vec{k}_0\neq\vec{0}$; more generally it will be useful for situations in which scalar fields are used as relational coordinates in space, since such scalars must necessarily be inhomogeneous.

At least in simple enough cases, one can derive an effective Friedmann equation for averaged quantities in an inhomogeneous geometry.
We work in the usual Hamiltonian ADM formalism for general relativity \cite{ADM08}, in which the geometry is characterised by a spatial metric $q_{ij}$ and its conjugate momentum $p^{ij}$ which is related to the extrinsic curvature.\footnote{We denote the inverse of the spatial metric $q_{ij}$ by $q^{ij}$. Indices are lowered and raised by contracting with the spatial metric and its inverse, respectively.
}
In addition, we have matter scalar fields $\chi^\alpha$ and conjugate momenta $(\pi_\chi)_\alpha$.
For simplicity we will only consider a single matter field.
Our ansatz for periodic inhomogeneity is
\begin{eqnarray}
  q_{ij}(t, \vec{x})
  &=&
  a(t)^2
  \delta_{ij}
  \left(
    1+\epsilon\, A(t)\cos(\vec{k}_0 \cdot \vec{x})
  \right)
  \mathcomma
  \\
  p_{ij}(t, \vec{x})
  &=&
  p(t)
  \delta_{ij}
  \left(
    1+\epsilon\, \upsilon\, A(t)\cos(\vec{k}_0 \cdot \vec{x})
  \right)
  \mathcomma
  \\
  \pi_\chi(t, \vec{x})
  &=&
  \Pi(t)
  \left(
    1+\epsilon\, \delta\pi(t)\cos(\vec{k}_0 \cdot \vec{x})
  \right)
  \mathperiod
\end{eqnarray}
We assume an isotropic geometry and a single direction in which inhomogeneity appears.
We can take this direction to be aligned with the $x$-axis so that $\vec{k}_0 = k_0 \vec{e}_{x}$.
$\epsilon$ is a small parameter used for a perturbative expansion, and $\upsilon\neq 1$ is needed to ensure that $q_{ij}$ and $p_{ij}$ are not proportional.
All functions of time are subject to the ADM constraints.

The spatial gradient of the scalar field $\chi$ is obtained from the diffeomorphism constraint
\begin{equation}
  \pi_\chi\partial_i\chi
  -
  2{p^k}_{i;k}
  =
  0
  \mathperiod
\end{equation}
If $\upsilon=1$ we would necessarily have ${p^k}_{i;k}=0$ and hence also $\partial_i\chi = 0$, forbidding a nontrivial scalar field configuration.
If $\upsilon\neq 1$ there is a single constraint which can be solved algebraically for $\partial_x\chi$.
We have that $\partial_x\chi=O(\epsilon)$; the exact form will play no role in what follows.

We also need to consider the Hamiltonian constraint
\begin{equation}
  \frac{16\pi G}{\sqrt{q}}
  \left(
    p_{ij}p^{ij}
    -
    \frac{1}{2}(p_i^i)^2
  \right)
  -
  \frac{\sqrt{q}}{16\pi G}
  \,
  \threecurvature
  +
  \frac{1}{2\sqrt{q}}
  \pi_\chi^2
  +
  \frac{\sqrt{q}}{2}
  q^{ij}
  \partial_i\chi
  \partial_j\chi
  =
  0
  \mathcomma
\end{equation}
where $q$ is the determinant of the spatial metric and $\threecurvature$ is the Ricci scalar of the spatial metric.
Expanded up to first order in $\epsilon$, this takes the form
\begin{equation}
 \begin{aligned}
  0
  &=
  -\frac{24\pi G p^2}{a^7}
  +
  \frac{\Pi^2}{2a^3}
  \\
  &
  \quad
  +
  \epsilon\cos(k_0 x)
  \left\{
    -\frac{24\pi G p^2}{a^7}
    \left(
      2\upsilon
      -
      \frac{7}{2}
    \right)
    A
    -
    \frac{a}{8\pi G}
    k_0^2
    A
    +
    \frac{\Pi^2}{2a^3}
    \left(
      2\delta\pi
      -
      \frac{3}{2}
      A
    \right)
  \right\}
 \end{aligned}
\end{equation}
resulting in a constraint at $O(\epsilon^0)$ and one relating the perturbations $A$ and $\delta\pi$ at $O(\epsilon)$.

The idea is now to define averaged quantities that incorporate inhomogeneities, and derive an effective Friedmann equation for those.
This would then be analogous to the equations for the global quantities we define in the GFT formalism, such as the total volume.
Such quantities are defined with respect to an averaging region, whose choice might lead to potential ambiguities in the formalism.
In our example, if we assume that the quantities to be averaged are inhomogeneous only in the $x$-direction and periodic with period $\frac{2\pi}{k_0}$, we will choose the averaging region to be a product over a single period in $x$ with an arbitrary ``cell'' in $y$ and $z$.
For a scalar density $f$, we then define the average as
\begin{equation}
  \frac{
    \int
    \intmeasure[3]{\vec{x}}
    f
  }{
    \int
    \intmeasure[3]{\vec{x}}
  }
  =
  \frac{k_0}{2 \pi}
  \int_0^{\frac{2\pi}{k_0}}
  \intmeasure{x}
  f
  \mathcomma
\end{equation}
which is a scalar with respect to volume-preserving diffeomorphisms in the spatial hypersurface.
Diffeomorphisms changing the coordinate volume $\int \intmeasure[3]{\vec{x}}$ rescale the value of such an average, which therefore cannot have any physical meaning in itself, but all dynamical relations between different such averages, such as those characterised by the Friedmann equation in cosmology, must be invariant under such rescalings.

The first such averaged quantity is then proportional to the total volume,
\begin{equation}
  V(t)
  =
  \frac{
    \int
    \intmeasure[3]{\vec{x}}
    \sqrt{q}
  }{
    \int
    \intmeasure[3]{\vec{x}}
  }
  =
  \frac{k_0}{2\pi}
  \int_0^{\frac{2\pi}{k_0}}
  \intmeasure{x}
  \sqrt{q}
  =
  a(t)^3
  \left(
    1
    +
    \frac{3}{16}
    \epsilon^2
    A(t)^2
    +
    O(\epsilon^4)
  \right)
\end{equation}
where we had to go to $O(\epsilon^2)$ to obtain a nontrivial contribution from the averaging over $x$.

The notion of a time derivative of the volume must be defined in a covariant way.
In a gauge in which the shift vector vanishes, the extrinsic curvature is $K_{ij} = \frac{1}{2N}\odv{q_{ij}}{t}$ and hence
\begin{equation}
  \frac{V'(t)}{N(t)}
  =
  \frac{k_0}{2\pi}
  \int_0^{\frac{2\pi}{k_0}}
  \intmeasure{x}
  \sqrt{q}
  K_i^i
  =
  -8\pi G
  \left(
    \frac{k_0}{2\pi}
    \int_0^{\frac{2\pi}{k_0}}
    \intmeasure{x}
    p_i^i
  \right)
  =
  -8\pi G
  T(t)
\end{equation}
where we used $p^{ij} = \frac{\sqrt{q}}{16\pi G} (K^{ij} - K^k_k q^{ij})$, $N(t)$ is the lapse and $T(t)$ is the average trace of the conjugate momentum:
\begin{equation}
  T(t)
  =
  \frac{k_0}{2\pi}
  \int_0^{\frac{2\pi}{k_0}}
  \intmeasure{x}
  p_i^i
  =
  \frac{3p(t)}{a(t)^2}
  \left(
    1
    +
    \frac{1-\upsilon}{2}
    \epsilon^2
    A(t)^2
    +O(\epsilon^4)
  \right)
  \mathperiod
\end{equation}
Finally, to characterise the scalar field we define its average energy
\begin{equation}
  E(t)
  =
  \frac{k_0}{2\pi}
  \int_0^{\frac{2\pi}{k_0}}
  \intmeasure{x}
  \frac{\pi_\chi^2}{2\sqrt{q}}
  =
  \frac{\Pi(t)^2}{2a(t)^3}
  \left(
    1
    +
    \epsilon^2
    \left(
      \frac{1}{2}
      \delta\pi(t)^2
      -
      \frac{3}{2}
      \delta\pi(t)
      A(t)
      +
      \frac{15}{16}
      A(t)^2
    \right)
  +
  O(\epsilon^4)
  \right)
  \mathperiod
\end{equation}
Using the two equations coming from the Hamiltonian constraint we then find
\begin{equation}
\label{eq:new_friedmann}
 \begin{aligned}
  &(8\pi G)^2
  \frac{T^2}{V}
  =
  24\pi G
  E
  \left[
    1
    +
    \left(
      -
      \left(
        \upsilon
        +
        \frac{1}{8}
      \right)
      A^2
      +
      \frac{3}{2}
      \delta\pi
      A
      -
      \frac{1}{2}
      \delta\pi^2
    \right)
    \epsilon^2
    +
    O(\epsilon^4)
  \right]
  \\
  &=
  24\pi G
  E
  \left[
    1
    +
    \left(
      -
      \frac{17}{8}
      +
      \frac{3}{2}
      \upsilon
      -
      \frac{\upsilon^2}{2}
      +
      \frac{k_0^2(5-2\upsilon)V^{1/3}}{32\pi G E}
      -
      \frac{2k_0^4V^{2/3}}{(32\pi G)^2 E^2}
    \right)
    \epsilon^2
    A^2
    +
    O(\epsilon^4)
  \right]
 \end{aligned}
\end{equation}
whose $O(\epsilon^0)$ part is the standard Friedmann equation of homogeneous cosmology,
\begin{equation}
  (8\pi G)^2
  \frac{T(t)^2}{V(t)}
  =
  \frac{V'(t)^2}{N(t)^2 V(t)}
  =
  24\pi G
  \rho(t)V(t)
  =
  24\pi G
  E(t)
  \mathperiod
\end{equation}
The periodic inhomogeneity we have introduced now leads to a correction to the right-hand side of the Friedmann equation \eqref{eq:new_friedmann} which depends on $\epsilon$, $\upsilon$, the perturbation $A(t)$, the wavenumber $k_0$ and the unperturbed variables. If we assume a regime where
\begin{equation}
  \frac{V(t)^{1/3}k_0^2}{32\pi G E(t)}
  \ll
  1
\end{equation}
and we choose $A(t)$ to be a constant $A_0$, then for given $\epsilon$, $\upsilon$ and $A_0$ the effective Friedmann equation would appear to be the one for an unperturbed universe but with an additional $O(1)$ multiplicative constant in front of the matter energy density, similar to what we have seen in the late-time limit of effective GFT cosmology in \cref{sec:late_times}.

\end{document}